\documentclass[sigconf]{acmart} 

\usepackage{kantlipsum,setspace,titlesec}
\usepackage{todonotes}

\usepackage{algorithm}
\usepackage[noend]{algorithmic}
\usepackage{url}
\usepackage{fancyvrb}
\usepackage{listings}
\usepackage{multirow}
\usepackage{booktabs}
\usepackage{caption}
\usepackage{amsmath}
\usepackage{subcaption}

\usepackage{tikz}
\usepackage{amsmath}
\usepackage{amsthm}

\usepackage{amssymb}
\usepackage{xspace}
\usepackage{xurl}
\usepackage{filecontents}
\usepackage{cleveref}
\usepackage{comment}
\usepackage{graphicx}
\usepackage[font={bf,it,footnotesize},labelsep=colon, belowskip=-1.0pt, aboveskip=0.3pt]{caption}
\usepackage[shortlabels]{enumitem}
\setlist[enumerate]{nosep}
\setlist{nolistsep,leftmargin=4.0mm}
\usepackage{mathtools}
\usepackage[hang,flushmargin]{footmisc} 
\usepackage{algorithm}
\usepackage[noend]{algorithmic}
\usepackage{kantlipsum,setspace,titlesec}
\usepackage{subcaption}
\usepackage{tikz}

\DeclareUnicodeCharacter{266B}{\textmusicalnote}

\usepackage{xpatch}
\newcommand{\sys}[0]{\textsc{Cache-Craft}\xspace}

\newcommand{\llama}[0]{\texttt{LLaMA}\xspace}
\newcommand{\vllm}[0]{\texttt{vLLM}\xspace}

\newcommand{\RPE}[0]{\texttt{RPE}\xspace}

\newcommand{\fullcache}[0]{\textsc{Full-Cache}\xspace}

\newcommand{\fullrecomp}[0]{\textsc{Full-Recomp}\xspace}

\newcommand{\ho}[0]{\textsc{Prefill-H2O}\xspace}

\newcommand{\randrecomp}[0]{\textsc{Random-Recomp}\xspace}

\newcommand{\prefixcache}[0]{\textsc{Prefix-Cache}\xspace}

\newcommand{\setcache}[0]{\textsc{Set-Cache}\xspace}

\newcommand{\LLMLingua}[0]{\textsc{Lingua2}\xspace}

\newcommand{\MapReduce}[0]{\textsc{MapReduce}\xspace}

\newcommand{\X}{{\sc Sys-X}\xspace}
\newcommand{\Y}{{\sc Sys-Y}\xspace}

\iftrue
\newcommand{\sm}[1]{\todo[inline,color=brown!40]{Subrata: #1}}
\newcommand{\sa}[1]{\todo[inline,color=green!40]{Shubham: #1}}
\newcommand{\re}[1]{\todo[inline,color=red!40]{Remark: #1}}
\newcommand{\rs}[1]{\todo[inline,color=blue!40]{Rounak: #1}}
\newcommand{\ag}[1]{\todo[inline,color=yellow!40]{Archit: #1}}

\definecolor{darkgreen}{rgb}{0.0, 0.6, 0.0}

\else
\newcommand{\sm}[1]{}
\newcommand{\sa}[1]{}
\newcommand{\re}[1]{}
\newcommand{\rs}[1]{}
\newcommand{\ag}[1]{}

\fi

\AtBeginDocument{%
  }

\renewcommand\footnotetextcopyrightpermission[1]{} 
\setcopyright{none} 

\acmConference[Preprint]{~}{ArXiv}{2025}
\acmYear{}            
\acmMonth{}           
\acmISBN{}            
\acmDOI{}             
\acmPrice{}           
\acmSubmissionID{}    
\acmBooktitle{}       

\usepackage{soul}
\usepackage{booktabs}
\usepackage{graphicx}         
\usepackage{tcolorbox}        

\definecolor{darkorange}{RGB}{204, 102, 0}
\renewcommand{\hl}[1]{{\textcolor{black}{#1}}}

\usepackage[shortlabels]{enumitem}
\setlist[enumerate]{nosep}
\setlist{nolistsep,leftmargin=4.0mm}
\setlist[itemize]{left=0pt, topsep=0.1pt, partopsep=0.1pt, parsep=0pt, itemsep=0pt}

\usepackage{mathtools}
\usepackage[hang,flushmargin]{footmisc} 

\usepackage{indentfirst}
\begin{document}



\title{\vspace{-1em}\sys: Managing Chunk-Caches for Efficient Retrieval-Augmented Generation
}


\author{\Large Shubham Agarwal$^1$\footnotemark[1], Sai Sundaresan$^1$\footnotemark[1], Subrata Mitra$^1$\footnotemark[2], Debabrata Mahapatra$^1$}
\author{\Large Archit Gupta$^2$\footnotemark[3], Rounak Sharma$^3$\footnotemark[3], Nirmal Joshua Kapu$^3$\footnotemark[3], Tong Yu$^1$, Shiv Saini$^1$
}

\affiliation{%
  \Large {$^1$Adobe Research} \hspace{20pt} 
  {$^2$IIT Bombay} \hspace{20pt} 
  {$^3$IIT Kanpur}
  \country{}
}

\renewcommand{\shortauthors}{Agarwal et al.}



\sethlcolor{yellow!40}

\begin{abstract}
Retrieval-Augmented Generation (RAG) is often used with Large Language Models (LLMs) to infuse domain knowledge or user-specific information. In RAG, given a user query, a retriever extracts chunks of relevant text from a knowledge base. These chunks are sent to an LLM as part of the input prompt. Typically, any given chunk is repeatedly retrieved across user questions. However, currently, for every question, attention-layers in LLMs fully compute the key values (KVs) repeatedly for the input chunks, as state-of-the-art methods cannot reuse KV-caches when chunks appear at arbitrary locations with arbitrary contexts. Naive reuse leads to output quality degradation.  This leads to potentially redundant computations on expensive GPUs and increases latency. In this work, we propose \sys, a system for managing and reusing precomputed KVs corresponding to the text chunks (we call \textit{chunk-caches}) in RAG-based systems. We present how to identify \hl{\textit{chunk-caches} that are reusable}, how to efficiently perform a small fraction of recomputation to \textit{fix} the cache to maintain output quality, and how to efficiently store and evict \textit{chunk-caches} in the hardware for maximizing reuse while masking any overheads. With real production workloads as well as synthetic datasets, we show that \sys reduces redundant computation by \textbf{51\%} over SOTA prefix-caching and \textbf{75\%} over full recomputation.
\hl{Additionally, with continuous batching on a real production workload, we get a \textbf{1.6$\times$} speedup in throughput and a \textbf{2$\times$} reduction in end-to-end response latency over prefix-caching while maintaining quality, for both the \llama-3-8B and \llama-3-70B models. 
}
\end{abstract}

\maketitle



\renewcommand{\thefootnote}{\fnsymbol{footnote}}
\footnotetext[1]{Equal contributions. \hspace{0.2em} \footnotemark[2] Corresponding Author (subrata.mitra@adobe.com).}
\footnotetext[3]{Work done at Adobe Research.}

\section{Introduction}
\label{sec:intro}

\begin{figure}[t]
     \centering
        \begin{subfigure}{0.48\linewidth}
         \centering
          \includegraphics[width=0.8\linewidth]{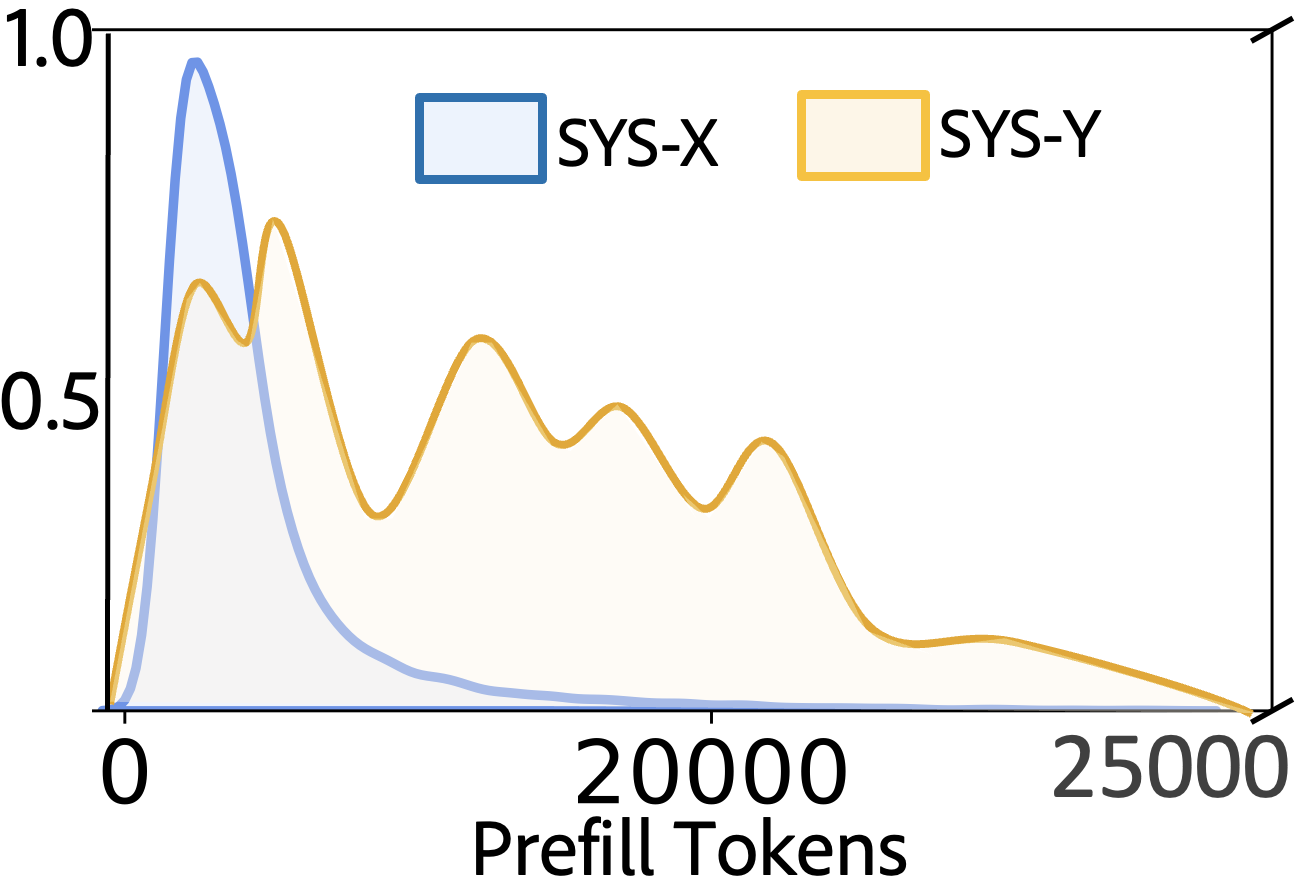}
         \end{subfigure}~
         \hfill
        \begin{subfigure}{0.48\linewidth}
          \centering
          \includegraphics[width=0.8\linewidth]{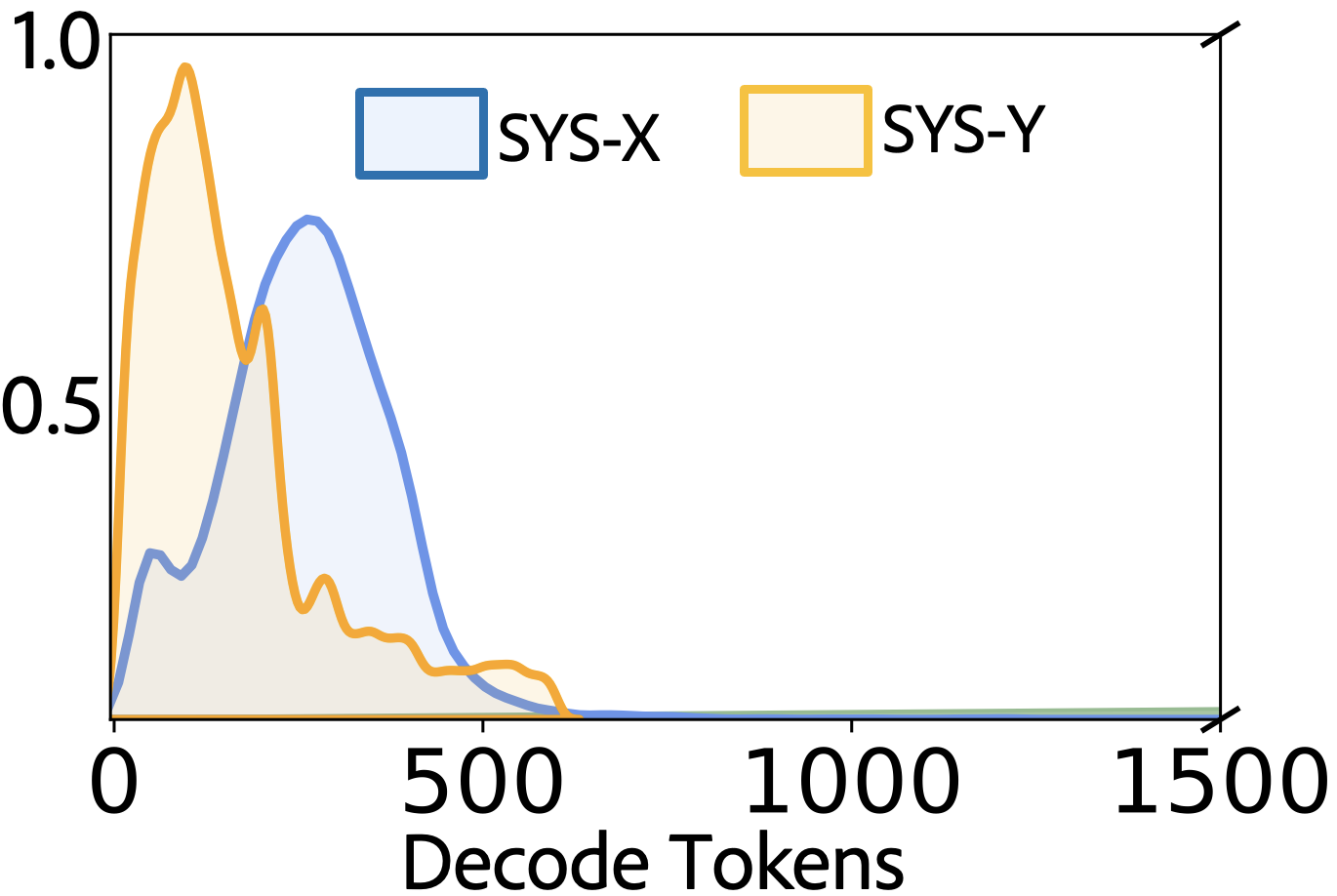}
       \end{subfigure}~
     \caption{Distribution of number tokens in prefill (left)  and decode (right phases for two real production RAG systems \X and \Y.}
     \label{fig:token_dist}
\end{figure}

\hl{Retrieval-Augmented Generation (RAG)} allows LLMs to access relevant context from a custom knowledge base outside its training data to generate grounded responses. 
RAG systems do not require model retraining and are therefore a cost-effective way to customize an LLM's output such that it is relevant and accurate with respect to a target \hl{knowledge base}.
The key components of a RAG-based system are a vector database and an LLM. The vector database stores the embedding of text chunks from a specific domain as indexes. 
During the \textit{retrieval} phase, relevant chunks are extracted based on these embeddings, using vector-similarity search~\cite{echihabi2021new, echihabi2020high}. 

In the \textit{generation} phase, the LLM uses the retrieved context to generate a response to \hl{the user's question}. The LLM processes its input prompt (retrieved chunks + user's question) in the \texttt{prefill} phase, building an initial computed-state called \textit{Key-Value} cache (KV-cache), which is then used in the \texttt{decode} phase for autoregressive token generation. The prefill phase is compute-bound because it processes all tokens of the input prompt in parallel; while the decode phase, which generates one token at a time, is memory-bound.

We aim to understand the bottlenecks in these systems. In this regard, we use two real production RAG systems, referred to as \X and \Y, for workload characterization, motivation, and evaluations.
\X helps users in setting up complex workflows for an enterprise SaaS product by answering queries and providing steps from user manuals and \Y helps users search and understand concepts from large \hl{knowledge bases} through repeated Q\&A.

\textbf{Computational bottlenecks in RAG-systems:}
Typically for RAG systems, the answers generated as well as user questions are short. However, longer input context with more relevant information is often crucial for the system to generate a well-informed answer~\cite{cuconasu2024power}. 
We highlight this in Fig. ~\ref{fig:token_dist} we show distributions of prefill (left) and decode (right) tokens for \X and \Y. It can be seen, that the number of prefill tokens is much more.

\begin{figure}[t]
    \centering
    \begin{minipage}{0.5\linewidth}
        \centering
        \includegraphics[width=\linewidth]{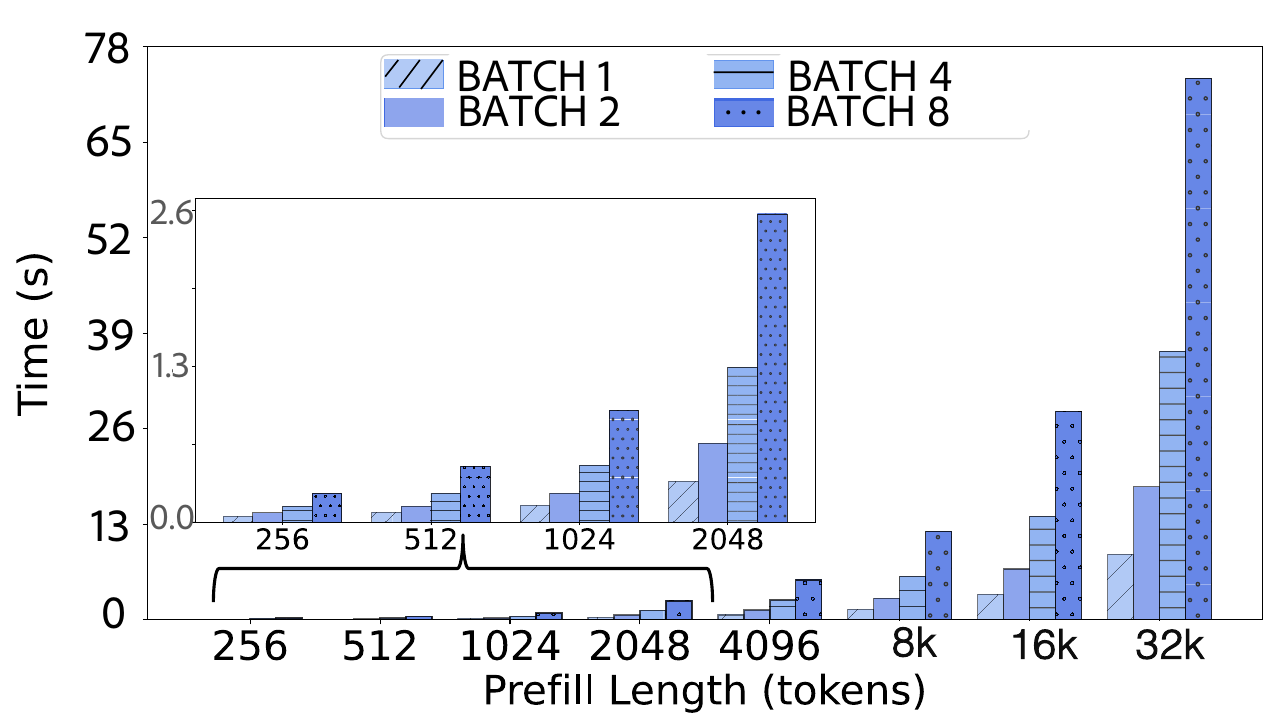}
        \caption{Prefill time across prefill length and batch size in \vllm on A100 80GB with TP=4.}
        \label{fig:prefill_times_into}
    \end{minipage}
    \hspace{0.5em}
    \begin{minipage}{0.4\linewidth}
        \centering
        \includegraphics[width=\linewidth]{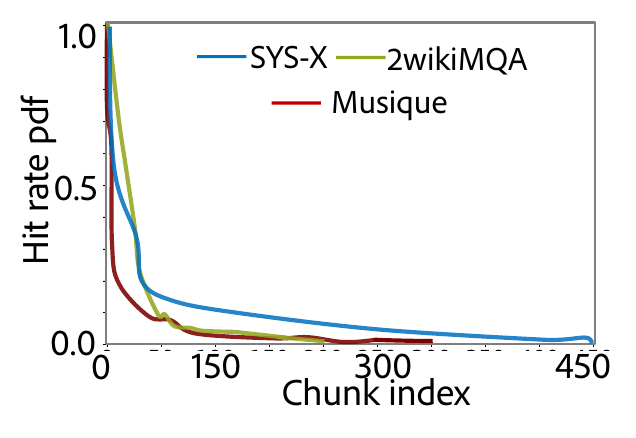}
        \caption{\textit{Chunk-cache} hit rate pdf for both \X and RAG datasets.}
        \label{fig:lmsys_pd}
    \end{minipage}
\end{figure}

\hl{The prefill time increases} quadratically with the length of input context, due to the attention computation in the transformer architecture \cite{vaswani2017attention}. 
{Fig.~\ref{fig:prefill_times_into}} shows prefill time increases with input token length across different batch sizes for \llama-3-70B~\cite{touvron2023llama} using \vllm~\cite{kwon2023efficient} with 4 NVIDIA A100-80GB GPUs, reaching up to \textbf{$\textbf{76 seconds}$} for $32k$-token sequences at a batch size $8$.
In real production workloads, the prefill time can often cross more than 100 seconds when serving multiple concurrent users. 
This increases the time-to-first-token (TTFT) ~\cite{agrawal2024taming} and degrades the user experience, as no response is generated until the whole input context is processed.
The impact of the prefill phase on overall latency is significant. In \X, it accounts for up to 77\% of total inference time. 

This problem is further exacerbated by the emergence of new LLMs, that can consume up to 1 million tokens (e.g., Claude 3~\cite{anthropic2024claude} and Gemini~\cite{reid2024gemini}). 
As more chunks can be used to improve response quality, longer context would lead to even longer TTFT.

\textbf{Opportunities for optimizing prefill in RAG:}
RAG systems typically operate on a finite knowledge base~\cite{lewis2020retrieval}. Moreover, our analysis, shown in Fig.~\ref{fig:lmsys_pd} for \X and RAG datasets like 2WikiMQA and MuSiQue, reveals that a subset of \hl{chunks gets retrieved frequently} by the system. For \X, 75\% of the retrieved chunks for a query were reprocessed, amounting to over 12B tokens in a month. Processing these tokens would require 9600 hours of GPU compute on \llama-3-70B using 8 A100 GPUs, costing approximately \$50k.

\textbf{Challenges in KV-cache reuse in RAG:}
Indiscriminate reuse of KV-caches from previously processed parts of the knowledge base can disrupt the relative positions of tokens, violating causal attention and degrading output quality~\cite{vaswani2017attention}. 
Additionally, for RAG systems with \hl{large} knowledge bases, precomputed KV-caches may not remain in GPU memory, as space is required for storing a) the LLM parameters and b) the growing KV-cache during the decode phase. The latency of loading precomputed KV-caches into GPU memory must not negate the savings from bypassing recomputation, requiring efficient system design and implementation.

\textbf{Limitations of existing works:}
Recent efforts to reduce prefill time and cost, such as \textit{Paged Attention}~\cite{kwon2023efficient}, \textit{CacheGen}~\cite{liu2023cachegen}, \textit{Radix Attention}~\cite{zheng2023efficiently}, \textit{RAG cache}~\cite{jin2024ragcache}
and \textit{context caching} in \textit{Gemini}~\cite{reid2024gemini} rely on \textit{prefix caching}, 
where different prompts with an identical prefix share KV-cache.
\hl{While this preserves the output quality by maintaining \textit{causal attention}~\cite{vaswani2017attention}, its usefulness is very limited in RAG systems because the RAG-system-retrieved text chunks and their relative ordering are sensitive to 
the input question. Slight variation in the user question can result in different sets of chunks and ordering, rendering the prefix caching technique ineffective. }
We found in production workloads, exact prefix caching applies to only a small fraction (8\%) of requests and 18\% of total prefill tokens.

\begin{figure}[t]
    \centering
    \includegraphics[width=0.8\columnwidth]{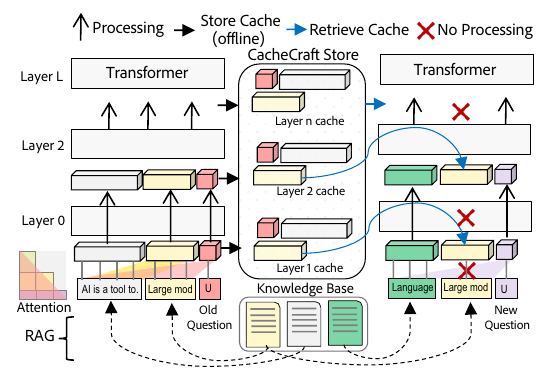}
    
    \caption{Overview of \sys}
    \label{fig:rag_overview}
\end{figure}

\textbf{\sys:}
We propose \sys, a system
for managing and reusing precomputed KV-caches in RAG. 
We overcome the challenges by (1) efficiently identifying which \textit{chunk-caches} can be reused even if their prefix alters, \hl{(2) identifying how to recompute the KV of a few selected tokens of the prefix-altered-caches to prevent quality degradation}, and (3) how to manage these caches such that most important chunks are prioritized, the overhead of load/store is masked to effectively reduce expensive GPU compute and TTFT latency for a workload. 
Fig.~\ref{fig:rag_overview} illustrates 
\sys:

\begin{itemize}
\item 
On the left, we show how KV-caches are formed across the \texttt{Transformer} layers when attention computation is done on the text chunks (shown with \textit{yellow} and \textit{gray}) corresponding to a question $Q$.
These pre-computed \textit{chunk-caches} are stored and managed by \sys along with some metadata. 

\item On the right we show their reuse. For a \textit{new question} two chunks of the knowledge-base become important. 
\sys identifies that it already has a cache for the \textit{yellow} chunk, therefore it retrieves and reuses the caches at the appropriate layers and only computation for the new \textit{green} chunk happens across layers. 

\item \hl{When a \textit{chunk-cache} is reused, KV is recomputed (not shown here) for a limited number of \textit{tokens} that were originally contextualized by tokens outside the chunk.
\sys further reduces this recomputation by using the relevance of a chunk w.r.t. the new question.
Tokens of less relevant chunks are not recomputed beyond a certain number of layers.}

\item \sys prioritizes storing KV-caches for the \hl{\textit{important} chunks} to maximize computation savings. The importance of a chunk is determined by its potential for direct reuse without significant recomputation, as well as its expected frequency of use based on the RAG's workload.

\end{itemize}

We implement \sys and integrate its KV-cache management capabilities into \vllm~\cite{kwon2023efficient}, a widely used package for LLM inference. The implementation is non-trivial
as it \hl{incorporates optimizations} such as \textit{FlashAttention}~\cite{dao2022flashattention} and \textit{PagedAttention}~\cite{kwon2023efficient} to enhance the \textit{Arithmetic Intensity}~\cite{ofenbeck2014applying} of computations.

\hl{We evaluate \sys with \llama in real deployment scenarios based on public traces. We show that it achieves a 51\% reduction in GPU computation costs for production workloads compared to prefix-caching (\S\ref{sec:eval_performance}). Under continuous batching through \textit{ORCA} for \X, \sys improves throughput by 1.6$\times$ and reduces end-to-end response latency by 2.1$\times$ for \llama-3-8B model and for \llama-3-70B, it provides a 1.6$\times$ speedup in throughput and 2$\times$ reduction in end-to-end response latency compared to prefix-caching (\S\ref{sec:deploy_performance_eval}). In both cases, 30\% tokens are recomputed which maintains 90\% of the base ROUGE F1 score on average.}

In summary, this paper makes the following contributions:
\begin{enumerate}

    \item We analyze real \hl{production workloads} to show that RAG systems are prefill-heavy, yet prefix caching remains ineffective.

    \item We present the key challenge of reuse, stemming from causal attention calculation through a formal problem formulation, and present detailed techniques to identify the reusability of \textit{chunk-caches} along with an efficient recomputation strategy to fix any potential \hl{degradation in generation quality}.  
    
    \item We present end-to-end design details and rationale for \sys, which is our optimized KV-cache management system for RAG, implemented in \vllm,
    a widely used LLM \hl{inference} package and plan to open-source it. 
    
    \item We present extensive evaluations on real-world, large production RAG systems, along with six other datasets, supported by a human evaluation user study and several sensitivity studies.
\end{enumerate}

\begin{figure*}[t]
    \centering
    \begin{minipage}{0.24\textwidth}
        \centering
        \includegraphics[width=1.0\linewidth]{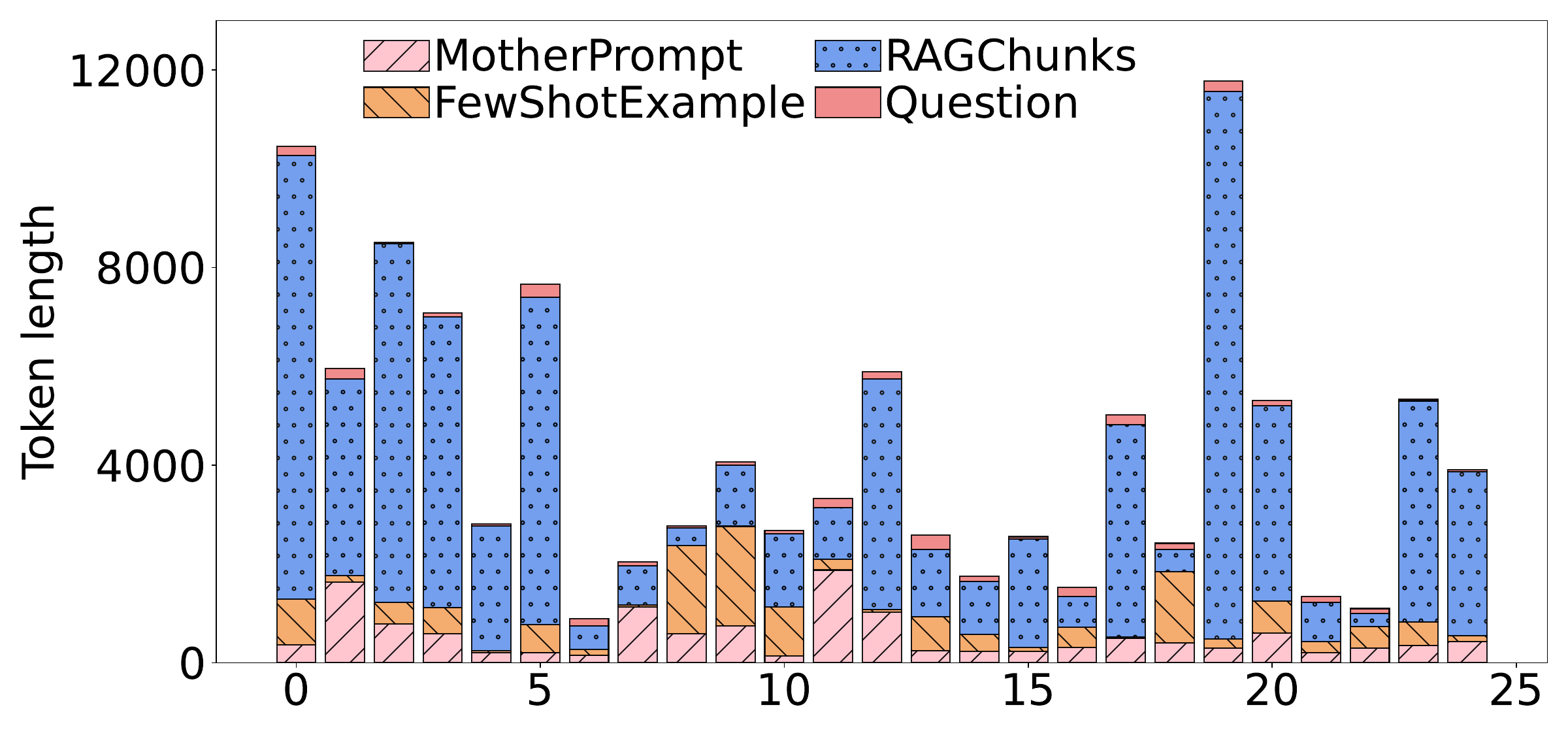}
        \caption{Token distribution of different prompt components (Mother prompt, RAG chunks, Examples, Query, etc.) across RAG use cases.}
        \label{fig:prompt_parts}
    \end{minipage}
    \hfill
    \begin{minipage}{0.5\textwidth}
        \centering
        \begin{subfigure}{0.32\textwidth}
            \centering
            \includegraphics[width=1.01\linewidth]{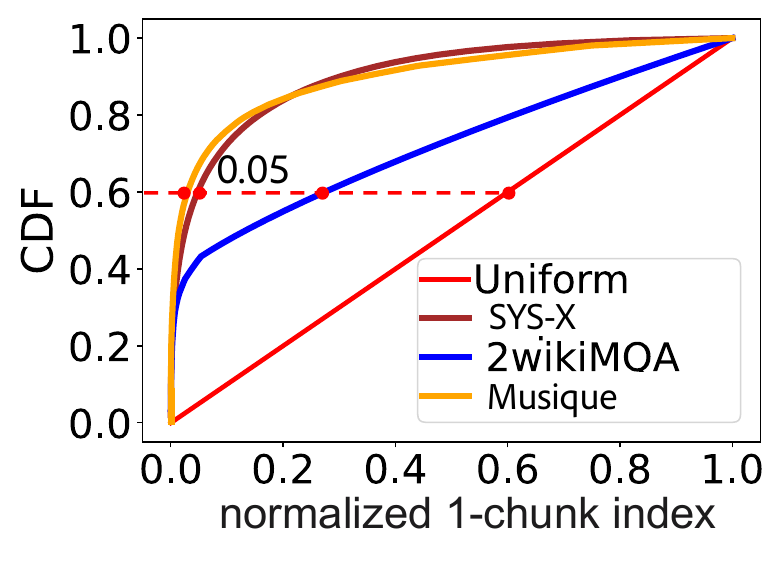}
            \vspace{-1.75em}
            \caption{Individual chunks}
            \label{fig:chunks_cdf}
        \end{subfigure}
        \hfill
        \begin{subfigure}{0.32\textwidth}
            \centering
            \includegraphics[width=1.01\linewidth]{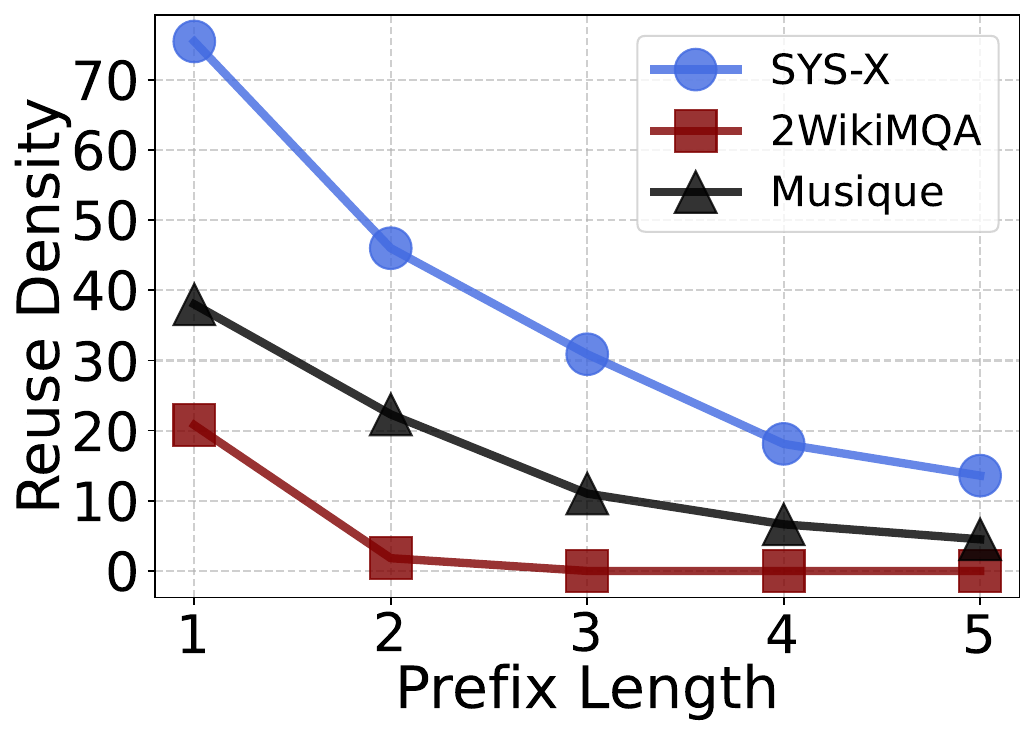}
            \vspace{-1.75em}
            \caption{Reuse Density}
            \label{fig:reuse_cdf}
        \end{subfigure}
        \hfill
        \begin{subfigure}{0.32\textwidth}
            \centering
            \includegraphics[width=1.01\linewidth]{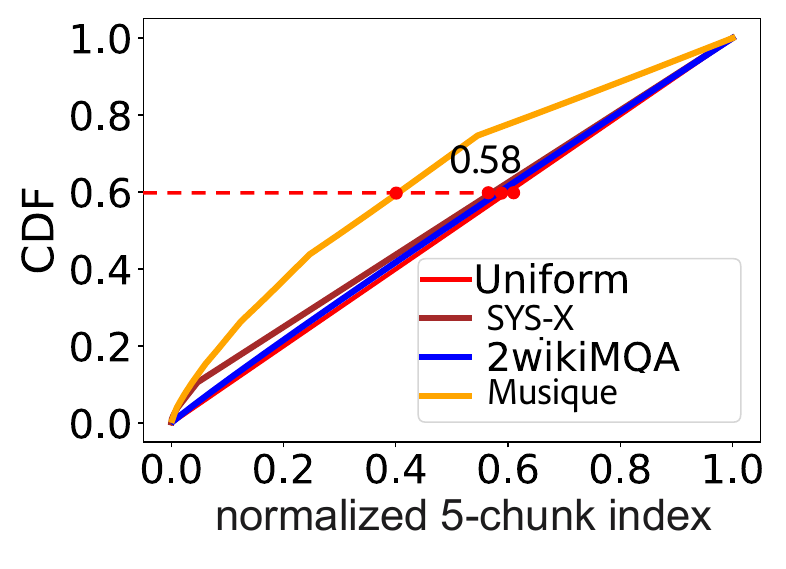}
            \vspace{-1.75em}
            \caption{Observed 5-tuples}
            \label{fig:prefix_cdf}
        \end{subfigure}
        \caption{Fig. \ref{fig:prefix_cache}(a) and \ref{fig:prefix_cache}(c) show the CDF of retrieval hit rates of the individual chunks and the observed $5-$tuple chunks respectively, across all user requests. Fig. \ref{fig:prefix_cache}(b) shows the decreasing cache reuse density with increasing prefix lengths.}
        \label{fig:prefix_cache}
    \end{minipage}
    \hfill
     \begin{minipage}{0.24\textwidth}
        \centering
        \includegraphics[width=1.0\linewidth]{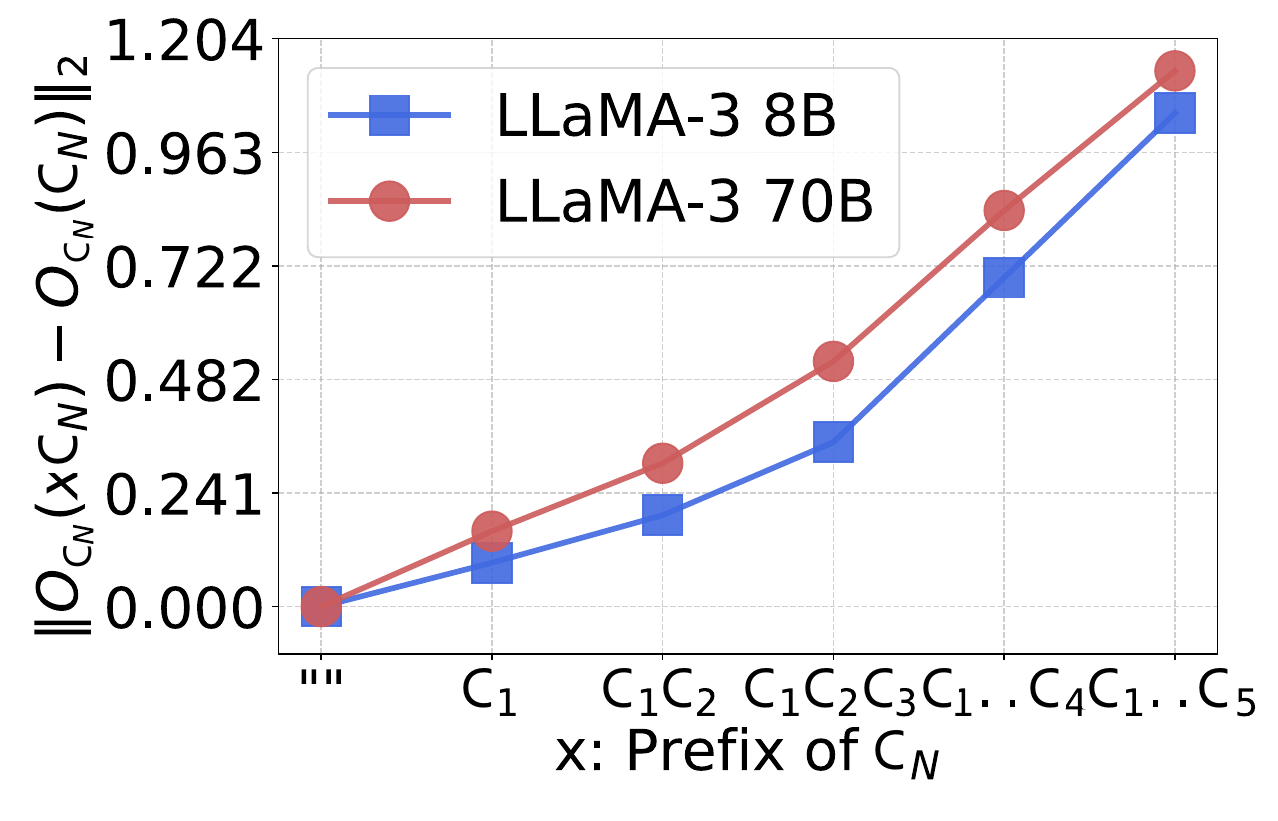}
        \caption{Deviation in output for chunk $C_1$ with increasing prefix chunks}
        \label{fig:prefix_diff_attn}
    \end{minipage}
\end{figure*}

\section{Background and Motivation}
\label{sec: background}

\subsection{\hl{Preliminaries of LLM}}
\label{sec: Preliminaries}
A transformer-based LLM progressively contextualizes a sequence of tokens $S=\{t_1, \cdots, t_n\}$ using $L$ \textit{transformer} layers. Each layer $l\in [L]$ receives $d-$dimensional embeddings of $n$ tokens, $H^l \in \mathbb{R}^{n\times d}$, as input, and outputs contextualized embeddings $H^{l+1} \in \mathbb{R}^{n\times d}$, which are known as hidden states. We denote the LLM operations, from input tokens $S$ all the way up to the last hidden states $H^L$, as $H^L(S)$.
The last layer hidden states are used in a task-specific manner. In the text generation task, the hidden embedding of the last token \hl{$H^L_{n}(S) \in \mathbb{R}^d$} is used to predict the $(n+1)^{\text{th}}$ token. 

In $l^{\text{th}}$ transformer layer, first, the $H^l$ is linearly transformed into the Query, Key, and Value matrices, $Q, K, V \in \mathbb{R}^{n\times d}$ respectively. 
The $Q$ and $K$ matrices are further transformed by positional embeddings (either absolute \cite{su2024roformer} or relative \cite{likhomanenko2021cape}) to capture the sequential order of the tokens.
Then the attention mechanism contextualizes the value embedding of $j^\text{th}$ token as $\tilde{V}_j = \mathrm{softmax}(Q_jK_{:j}^T) V_{:j}$, where $Q_j \in \mathbb{R}^{1\times d}$ is the $j^\text{th}$ query vector and $K_{:j}, V_{:j} \in \mathbb{R}^{j\times d}$ are all the key and value vectors up to the $j^\text{th}$ token. Finally, the contextualized hidden state $H^{l+1}_j$ is obtained by normalizing $H^l_j +  \mathrm{FNN}(\tilde{V}_j)$. 

\hl{LLM operates in two phases. In the \textbf{prefill} phase, it contextualizes all available prompt tokens. The hidden states $H^L(S)$ are computed for the prompt $S=\{t_1, \cdots, t_n\}$ using the matrix operation}
\begin{align}\label{eq:attn}
    \tilde{V} = \mathrm{softmax}(QK^T \odot M)V,
\end{align}
where $\odot$ denotes element-wise product, and $M \in \{0,1\}^{n\times n}$ is a lower triangular matrix, known as causal attention mask, to ensure each token attends only to its previous tokens. This attention computation is $O(n^2)$, \hl{as both $Q$ and $K$ matrices are of size $n\times d$.
 During this phase, the model generates the $KV$ pairs for all tokens in the sequence, which are used to predict the next token $t_{n+1}$.}
 
\hl{In the \textbf{decode} phase, the model generates tokens autoregressively. 
For each newly generated token $t_j$ starting from the position $j=n+1$, the attention mechanism is applied to contextualize its raw embedding $H^0_j$.
However, instead of recomputing the $KV$ for all previous tokens again, the model uses the cached $K$ and $V$ matrices (KV cache) from the prefill phase. By reusing these cached representations of the previous $n$ tokens at every layer, the computation is reduced from $O(n^2)$ to $O(n)$. As each new token is generated, the KV cache is updated by adding the new token's key and value.}

\hl{\textbf{KV cache vs. Prefix cache:} While KV cache optimizes the decode phase by reusing KV pairs of previously processed tokens, the prefill phase still requires $O(n^2)$ computation to establish the full context. Prefix-cache stores and reuses previously computed context that matches with the prefix of the input prompt, and only computes the rest~\cite{zheng2023efficiently} to reduce prefill computation. 
}

\subsection{Prefill Dominates Decode in RAG}\label{sec:prefill_vs_decode}
In a typical RAG-system, the overall prompt sequence $S$ consists of a few initial instruction text chunks, several retrieved chunks from a knowledge base, \hl{and the user's question or request $U$, i.e., ${S= C_1:C_kU}$, where $C_1:C_k$ denotes the concatenation of $k$ chunks.}\footnote{The instructions in the prompt are the same across all prompts. These instructions are similar to an always repeated chunk and can be dealt with under the same framework.}
\hl{In most production RAG systems, between 5 to 15 chunks are retrieved to answer a query $U$.} 
The overall length of prefill tokens $|S|$ and the lengths of their constituents may vary for different requests. 
We analyze this in Fig.~\ref{fig:prompt_parts} for a proprietary system, \X, from 25 sessions. The majority of the tokens (60\% to 98\%) are from the retrieved chunks from a knowledge base (in blue). A few tokens are from the mother prompt (instructions for the chatbot), few-shot examples (for in-context learning~\cite{dong2022survey}), and the user's questions.  

Due to the extra chunks apart from the user's question, the number of prefill tokens becomes significantly more than that of decode. To verify this, we analyze three systems: proprietary production \X and \Y, and another open-source LMSys~\cite{zheng2023lmsys} chat system. 
Fig.~\ref{fig:token_dist} shows a disparity in the number of tokens between prefill and decode: \textbf{30k} prefill tokens for \textbf{600} decode tokens on average.

We compare prefill times and operations on 4xA100-80GB GPUs using the \llama-70B. For \X, prefill accounts for 55.4\% of total time and 19.3x decode operations. \Y takes 76\% of the time and 46x the operations, while LMSys uses 22\% time and 4.4x operations. 

\textit{Contrary to the popular belief that decode is slow in LLMs, for RAG-systems prefill phase typically dominates both the amount of token computation and total latency, despite being highly parallelized.}

\subsection{Evidences of Chunk-Reuse}\label{sec:chunk_reuse}
Since prefill is the primary bottleneck in RAG, we find improvement opportunities by observing repetitions in chunk retrieval. If $N$ is the total number of chunks representing the knowledge base accessible for RAG, a significant portion of $N$ is retrieved multiple times across different user sessions, where a session consists of multiple user requests and LLM responses.

\hl{We substantiate this by analyzing the \textit{retrieval hit rates}, defined as the fraction of all retrievals (across multiple sessions) in which a particular chunk is present. Fig \ref{fig:chunks_cdf} shows the retrieval hit rates of 3 RAG systems:} \X, 2wikiMQA~\cite{ho-etal-2020-constructing} and MuSiQue~\cite{10.1162/tacl_a_00475}. The top 5\% of chunks are accessed by \textbf{60\%} of the requests in both \X and MuSiQue, and \textbf{40\%} requests in 2wikiMQA. In \X, most chunk reuse occurs across users (94\%), with reuse within a session at 55\% and across sessions at 67\%.
Exploiting the high reuse of knowledge chunks can optimize the prefill by reusing caches that are computed in the previous sessions, instead of recomputing for every retrieval. 
However, cache reuse across different user requests is non-trivial.

\subsection{Why Cache-Reuse is non-trivial?}
\label{sec:challenges}

\hl{\textbf{Limitations of Prefix-Caching:}} The prefix-cache approach is to store the KV-caches of ordered $k$-tuple chunks when they are co-retrieved for a request $U$, and reuse it for a future request $U'$ if the same $k$ chunks are retrieved in the same order. However, \hl{the \textit{reuse density}, defined as the number of ordered $k-$tuple chunks observed in previous requests}, drops significantly w.r.t. $k$, reducing the reusability of their cache. 
We analyze reuse density for 3 datasets over the most recent $1000$ requests in Fig.~\ref{fig:prefix_cache}b, and find that it drops 
to as low as 5 for $k=5$. 
This worsens further as we analyze the $5-$tuple retrieval hit rates, defined as the fraction of all observed $5-$tuple retrievals in which a particular $5-$ tuple is present. Fig.~\ref{fig:prefix_cache}c shows this hit rate is significantly low compared to those of the individual chunks in Fig.~\ref{fig:prefix_cache}a.
\hl{However, a high hit rate is crucial for cache utilization. Moreover, unlike Fig.~\ref{fig:prefix_cache}a, the distribution in Fig.~\ref{fig:prefix_cache}c does not follow the power law, indicating that, for a high re-usability, several k-tuple chunks should be cached. Therefore,
the combinatorial growth of the number of possible k-tuple makes the memory footprint of prefix-caching prohibitive.}

\hl{On the other hand, a method that can reuse the KV-cache of individual chunks (pre-computed while serving a past request) at any position, without restricting to the prefix order, would have a significantly high retrieval hit rate as evidenced in Fig. \ref{fig:prefix_cache}a.}

\begin{figure}[t]
    \centering
    \captionsetup[subfigure]{justification=centering}
    \begin{subfigure}[t]{0.48\linewidth}
        \centering
        \includegraphics[width=0.9\linewidth]{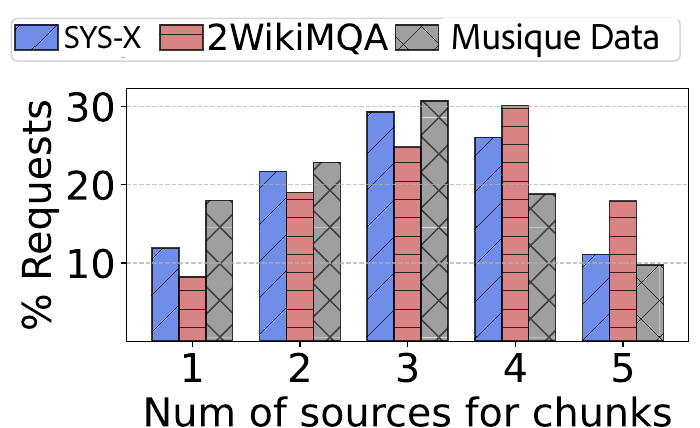}
        \caption{}
        \label{fig:chunk_sources}
    \end{subfigure}
    \hfill
    \begin{subfigure}[t]{0.48\linewidth}
        \centering
        \includegraphics[width=0.9\linewidth]{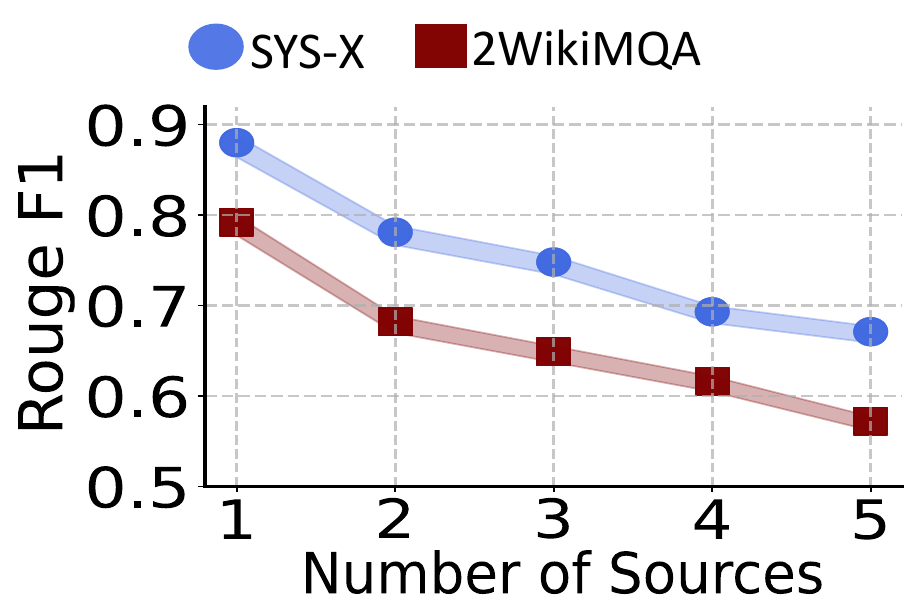}
        \caption{}
        \label{fig:motiv_shuffle}
    \end{subfigure}
    \caption{(a) Shows the distribution of chunk sources for non-prefix chunk reuse, while (b) Shows the impact of chunk reuse from multiple sources on ROUGE F1 scores, even when using new positional embeddings.}
    \label{fig:combined_plots}
\end{figure}

Although a high hit rate is encouraging, it presents a few major decision challenges. We lay them out in the following.

\hl{\textbf{Contextualization:}}
A chunk $C$ may have more than one stored KV-caches that were computed while serving different user requests in the past, e.g., $S^1=C_1^1\!:\!C_i\!:\!C_k^1U^1$ and $S^2=C_1^2\!:\!C_j\!:\!C_k^2U^2$, where $C=C_i=C_j$, but the positions $i$ and $j$ are not necessarily the same. Which of these KV caches of $C$ should be used for a new request $U^3$? 
\hl{A key challenge with reusing pre-computed KV-cache is contextualization. 
The stored KV-cache of $C_{i}$ have been contextualized by $C_1^1\!:\!C_{i-1}^1$. We analyze how the contextualization of $C_{i}$ changes with varying numbers of prefix chunks. In particular, we use the last layer hidden states $H^L_{C_i}(C_1\!:\!C_i) \in \mathbb{R}^{|C_i|\times d}$ corresponding to the tokens in $C_i$, when $C_1\!:\!C_i$ is given as input. Fig.~\ref{fig:prefix_diff_attn} shows its difference from that of no contextualization $H^L(C_i)$. Evidently, the contextualization grows with more prefix chunks.}

\hl{\textbf{Sensitivity to chunk ordering:} }
The relative ordering of prefix chunks affects the contextualization due to two reasons: \hl{a) the unidirectional attention by the causal attention mask $M$ in \eqref{eq:attn} and b) the positional embedding that alters $Q$ and $K$ matrices specific to the token positions.} More precisely, $H^L_{C_i}(C_1\!:\!C_i) \neq H^L_{C_i}\big(C_{(1)}\!:\!C_{(i-1)}C_i\big)$, where $C_{(1)}\!:C_{(i-1)}$ is a permutation of the prefix chunks.

\hl{\textbf{Cache from multiple sources:}} Another decision challenge occurs when the KV-caches are stored at different requests. Let $C$ and $C'$ are retrieved for serving $U$, the KV-cache of $C$ was stored from a past prompt $S_1=C_1^1\!:\!C\!:\!C_k^1U^1$ and that of $C'$ was stored from a different prompt $S_2=C_1^2\!:\!C'\!:\!C_k^2U^2$.
In such cases, can any of the chunk's KV-cache be used reliably to serve the new request $U$? Can both be used?
Fig.~\ref{fig:chunk_sources} shows that for a majority of the requests in \X and {\color{black}2wikiMQA}, the $5$ retrieved chunks were found in the retrievals of $3$ past requests. Finding all the $5$ chunks in the retrievals of only $1$ past request is not common (around 10\% for \X and {\color{black}8\% for 2wikiMQA}).
A naive reuse of KV-caches that were precomputed across different requests significantly degrades output quality. Our findings in Fig. \ref{fig:combined_plots}\subref{fig:motiv_shuffle} show a \textbf{50\%} drop in F1 score when all $5$ chunks are reused from five distinct past requests (Fig. \ref{fig:motiv_shuffle}), highlighting the need for a more advanced reuse strategy. 

\begin{figure}[t]
    \centering
    \begin{minipage}[t]{1.0\linewidth}
        \centering
        \includegraphics[width=1.0\linewidth]{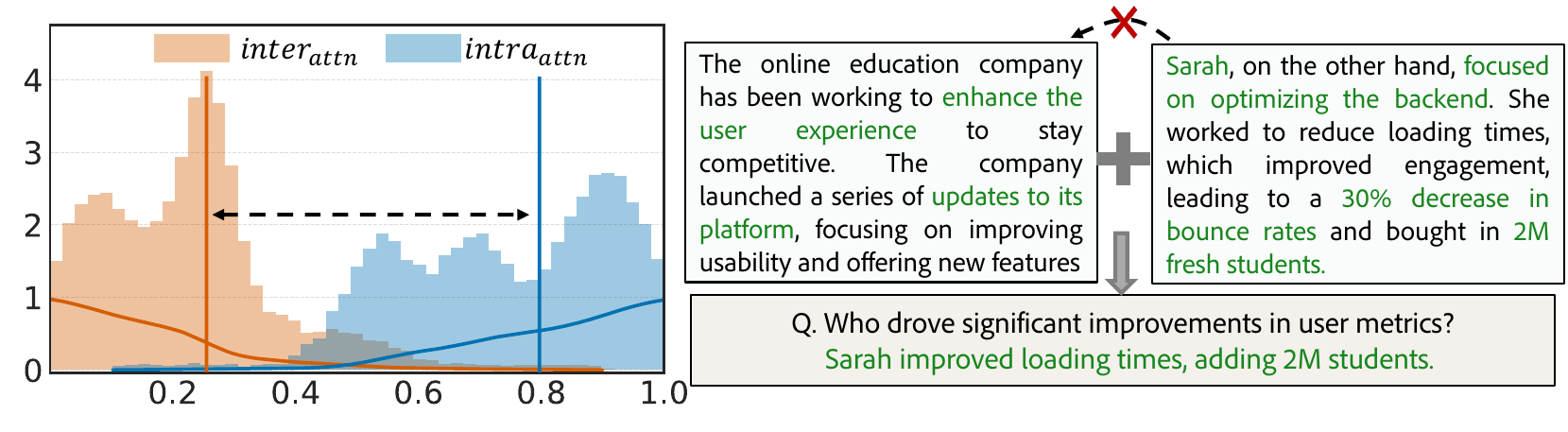}
        \caption{Inter-attention (C1, C2) and intra-attention (C2, C2) distributions for chunks C2 with C1 in context. The overlap in the distribution is less, meaning inter$<$intra, and hence the output for <C1, C2, Q> without letting C2 attend to C1 is correct due to less overlap indicating little contextualization.}
        \label{fig:tasks_quality_eval_}
    \end{minipage}
\end{figure}

\begin{figure}[t]
    \centering
    \begin{minipage}[t]{1.0\linewidth}
        \centering
        \includegraphics[width=1.0\linewidth]{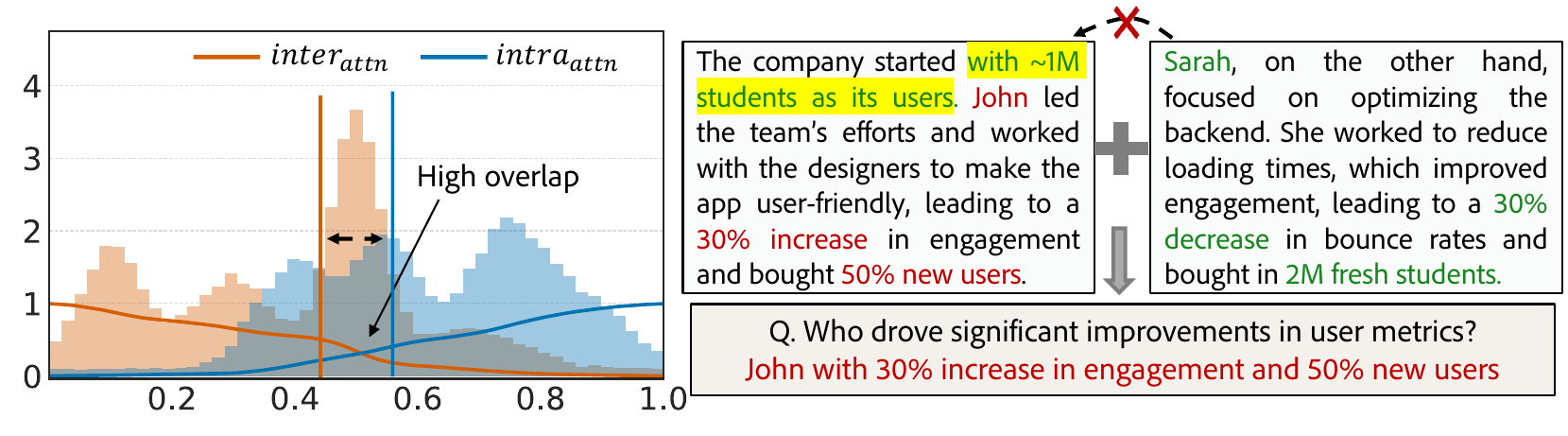}
        \caption{Inter-attention (C1, C2) and intra-attention (C2, C2) distributions for chunks C2 with C1 in context. The overlap in the distribution is more, meaning inter$\nless$intra, and hence the output for <C1, C2, Q> without letting C2 attend to C1 is incorrect due to more overlap indicating contextualization.}
        \label{fig:tasks_quality_eval2}
    \end{minipage}
\end{figure}

To understand when naive reuse of the KV-cache works and when it does not, 
we analyze two example prompts, 
and their outputs in Figs. \ref{fig:tasks_quality_eval_} and \ref{fig:tasks_quality_eval2}. We use $k=2$ relevant chunks $C_1,C_2$ to construct the prompt of a question $U$. The KV-caches of $C_1$ and $C_2$ are precomputed from $H^L(C_0C_1)$ and $H^L(C_0'C_2)$ respectively. We observe in Fig. \ref{fig:tasks_quality_eval2} that when the values of intra-chunk attention weights (from $Q_{C_2}K^T_{C_2}$) and inter-chunk attention weights (from $Q_{C_2}K^T_{C_1}$) are highly overlapping, naive reuse of stale KV-cache results in a wrong output. Whereas if they are less overlapping, the precomputed KV-cache can lead to the right answer in Fig. \ref{fig:tasks_quality_eval_}.

\section{\sys Design}
\label{sec:overview}

At a high level, \sys enhances a RAG system by managing the KV-caches of knowledge chunks, as illustrated in Fig.\ref{fig:rag_overview}. We denote the \textit{Chunk-Cache} of a chunk $C$ that was originally computed from $H^L(C_1\!:\!C_i\!:\!C_kU)$, while serving a request $U$ at the $i^\text{th}$ position (i.e., $C_i=C$), as
\begin{align}
\small
    \mathscr{C}\big(C \ |\ C_1\!:\!C_{i-1}\big) := \left\{\left(K_C^l, V_C^l\right)\ \middle| \ l\in [L]\right\}, 
\end{align} 
where $K_C^l$ and $V_C^l$ are the key and value vectors in $l^\text{th}$ layer corresponding to the tokens in $C$. 

Apart from storing the $\mathscr{C}\big(C \ |\ C_1\!:\!C_{i-1}\big)$, We also store certain metadata to determine whether a particular of KV-cache $C$ can serve a new request $U'$ in future. \sys operates in two phases: online and offline. 
The metadata computation is performed in the offline phase, and the determination of its ``usefulness'' is performed in the online phase, while serving a new request $U'$. 

In its online phase, \sys first selects the most useful (w.r.t. $U'$) version of \textit{chunk-cache} of $C$ out of all the stored versions $\mathscr{C}(C | \cdots)$. 
Then \sys selectively recomputes the key and value vectors for a few tokens of $C$ to contextualize w.r.t. $U'$. Clearly, if there are no \textit{chunk-caches} of $C$, then the key and value vectors for all tokens have to be computed afresh. Once the $K$ and $V$ matrices of $C$ are contextualized for $U'$, either by fixing a stored chunk or by computing afresh, \sys repeats this same online procedure for the chunk next to $C$ that is inline to serve $U'$.

\subsection{Determining Cache Reusability} 
\label{sec: cache_quality}

From our analysis in Fig. \ref{fig:tasks_quality_eval_} and \ref{fig:tasks_quality_eval2}, we observe that reusability can be assessed by determining how much a chunk’s KV computation is influenced by external context (tokens outside the chunk) versus its own tokens. If a chunk is mainly influenced by its own tokens, it is more likely to produce high-quality answers when reused.

In fact, a chunk with more tokens is more reusable because tokens closer to each other have stronger attention due to positional embeddings, compared to distant tokens from other chunks~\cite{su2024roformer}. 
To capture the attention within and across the chunks, we define the following two attention-based metrics:

\begin{enumerate}
     \item 
     \textbf{Inter attention} measures the cumulative attention weight from tokens in chunk $C_{i}$ to tokens in chunk $C_{j}$ where $i<j$:
    \begin{align}\label{eq:inter}
        inter(C_{i}, C_{j}) = \sum_{k \in C_{i}} \sum_{l \in C_{j}} a_{kl},
    \end{align}
    \hl{where $a_{kl}$ is the \textit{attention weight} from the $k^{th}$ token of chunk $i$ to the $l^{th}$ token of chunk $j$, computed from the softmax in \eqref{eq:attn}.}
    
    \item 
    \textbf{Intra attention} measures the cumulative attention weight within chunk $C_{i}$ from each token to previous tokens in the same chunk:   
    \begin{align}\label{eq:intra}
    \small
        intra(C_{i}) = \sum_{k,l\in C_{i}: k < l} a_{kl}.
    \end{align}
\end{enumerate}

The attention weights involved in the $inter$ and $intra$ are used to obtain the output from the attention computation. For instance, in case of $3$ chunks \( [C1, C2, C3] \), the attention output is
{\small
\begin{align}
\begin{bmatrix}
\tilde{V}_{C_1} \\
\tilde{V}_{C_2} \\
\tilde{V}_{C_3}
\end{bmatrix} =
\begin{bmatrix}
\small
\overline{intra}(C_1) & \mathbf{0} & \mathbf{0} \\
\overline{inter}(C_1, C_2) & \overline{intra}(C_2) & \mathbf{0} \\
\overline{inter}(C_1, C_3) & \overline{inter}(C_2, C_3) & \overline{intra}(C_3)
\end{bmatrix}
\begin{bmatrix}
V_{C_1} \\
V_{C_2} \\
V_{C_3}
\end{bmatrix},
\end{align}
}
where $\overline{intra}$ and $\overline{inter}$ represent the associated attention weights in \eqref{eq:inter} and \eqref{eq:intra} without summing them, \hl{$V_C$ represents the pre-attention value vectors of all tokens in chunk $C$, and $\tilde{V}_C$ represents the corresponding post-attention value vectors.}

Reusability of the cache $\mathscr{C}(C_3 | C_1C_2)$ for a new request depends on the new prefixes.
Consider $2$ cases with prefixes (i) $C_{4}$-$C_{2}$-$C_{3}$ and (ii) $C_{5}$-$C_{6}$-$C_{3}$. The first sequence carries the $C_2$ as a prefix similar to that of $\mathscr{C}(C_3 | C_1C_2)$, making it more reusable than the second sequence, which does not have any common chunk in its prefix.

Assuming that the higher the prefix overlap, the higher will be the reusability,
we calculate a 
\textit{Prefix Overlap Score $\beta$} for a \textit{chunk-cache} of $C_i$ corresponding to the current prompt sequence $S_{new}$ as:
\begin{align}
    \small
\beta(C_i \mid S_{new}) = \frac{\sum_{j \in S_{old} \cap S_{new}} inter(i,j)}{\sum_{j \in S_{old}} inter(i,j)},
\end{align}
where $S_{old}$ is the set of chunks forming $C_i$'s old prefix.

However, since $\beta$ simply sums the $inter$ attention terms for overlapping chunks, it is order-invariant and only captures the subset match between the previous and current prefixes. 
For instance, consider the two scenarios: \(C_{1}\)-\(C_{2}\)-\(C_{3}\) and \(C_{2}\)-\(C_{1}\)-\(C_{3}\). 
In both cases, $\beta$ equals 1, yet the potential for reusing the cached chunk \(C_{3}\) can differ significantly due to the reordering of the prefix sequence. 
Note, it is not prudent to manipulate the retrieved order of chunks to match the prefix of the cached-chunk, due to \textit{lost-in-the middle} phenomena with LLM-based RAG systems~\cite{liu2024lost}.

\begin{figure}[t]
    \centering
    \includegraphics[width=0.9\columnwidth]{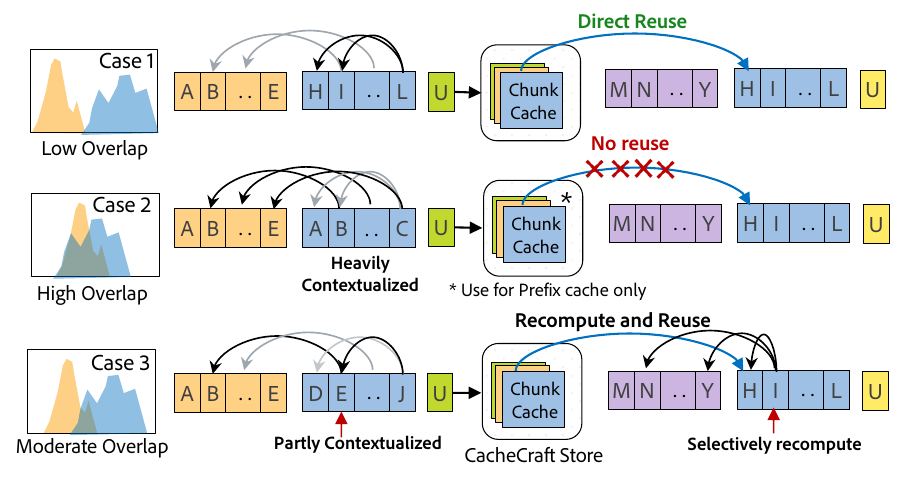}
    \caption{\textit{Chunk-cache} reuse scenarios. Inter and intra-attention for the blue chunk are shown on the left. Dark arrows represent high contextualization, and gray arrows indicate low. Case 1: Blue chunk is self-contextualized, so cache can be used even with new context with purple chunk. Case 2: Blue chunk is heavily contextualized on outside orange chunk, no reuse. Case 3: Only few tokens of blue are contextualized outside, so can be reused with selective recomputation.
    }
    \label{fig:contexual_cases}
\end{figure}

To account for prefix reordering, we introduce the \textit{Order Penalty Score ($\gamma$)}, which penalizes a chunk for different ordering in the prefix sequence. 
Let \( A_{old} = \langle C_i \mid C_i \in S_{old} \cap S_{new} \rangle \) 
denote the ordered sequence of chunks according to $S_{old}$'s order, and similarly $A_{new}$. We define $\gamma$ for chunk $C_{i}$ w.r.t. $S_{new}$ as the normalized Kendall's Tau distance \cite{cicirello2019kendall} between vector $A_{old}$ and $A_{new}$:
\begin{align}
\gamma(C_{i} \mid S_{new}) = \frac{D}{T},  \quad T = \binom{m}{2} = \frac{m(m-1)}{2},
\end{align}
where $m=|S_{old} \cap S_{new}|$ and \(D\) is the number of discordant pairs between \(A_{1}\) and \(A_{2}\). A higher value of \(D\) indicates a greater discrepancy in ordering, leading to a higher penalty for reuse. Hence, we adjust \(\beta\) to account for this discrepancy by penalizing it, resulting in the
\textit{Adjusted Prefix Overlap Score} denoted by
\begin{align}
    \beta^{'} (C_{i} \mid S_{new}) = \beta (C_{i} \mid S_{new}) \cdot (1 - \gamma (C_{i} \mid S_{new})).
\end{align}

Finally, to assess the reusability of a chunk across different prefix contexts, we measure how much the chunk's KV is contextualized by its prefix. A chunk’s KV is more reusable if it is \hl{a) less influenced by its prefix and b) more influenced by its own tokens.} 
We formulate these two effects by calculating as:
\begin{align}
    a(C_i) = \sum_{j < i} \frac{{inter}(C_j,C_i)}{|C_i| \cdot |C_j|} \quad \text{and} \quad b(C_i) = \frac{{intra}(C_i)}{|C_i|^2},
\end{align}
where $a$ is the normalized sum of $inter$-attention scores between chunk $C_i$ and its prefix chunks at the time of caching,
and $b$ is the normalized $intra$-attention score of chunk $C_i$. 
Normalizing w.r.t. the chunk length $|C|$ ensures comparability across chunks of varying sizes.
\hl{The layer-wise inter and intra values
are averaged to 
\begin{align}
    \Bar{a}(C_i) = \frac{1}{L}\sum_{l=1}^L a_l(C_i) \quad \text{and} \quad \Bar{b}(C_i) = \frac{1}{L}\sum_{l=1}^L b_l(C_i).
\end{align}}
\textit{A higher $\frac{\Bar{a}}{\Bar{b}}$ ratio indicates greater outside contextual influence on the chunk's KV.} 
We use this ratio to define the \textit{Cache Context Impact (CCI)} for chunk $C_i$ as
\begin{align}
    CCI(C_i) = \frac{1}{1 + e^{-\frac{\Bar{a}}{\Bar{b}}}},
\end{align}
\hl{where the \texttt{sigmod} function standardizes its range between $0$ and $1$.}
A high value of the $CCI$ for a cache indicates that the chunk is highly contextualized, reducing its potential for reuse unless the prefix context matches closely. Conversely, a low $CCI$ suggests that the chunk is largely independent of its prefix context, making it more reusable across different contexts. 
Fig. \ref{fig:contexual_cases} shows three different scenarios to illustrate how \textit{chunk-cache} is reused by \sys.
Inter- and intra-attention for the blue chunk are shown on the left and dark/black arrows for high contextualization, while gray arrows are for low contextualization. \textbf{Case 1:} The blue chunk is self-contextualized, allowing cache reuse even with the new purple chunk context. \textbf{Case 2:} The blue chunk is highly contextualized by the orange chunk, so no reuse is possible. \textbf{Case 3:} Only a few tokens in the blue chunk are contextualized externally, allowing partial reuse with selective recomputation.

\begin{figure}[t!]
    \centering
    \begin{minipage}[h!]{0.485\linewidth}
        \centering
        \includegraphics[width=1.0\linewidth]{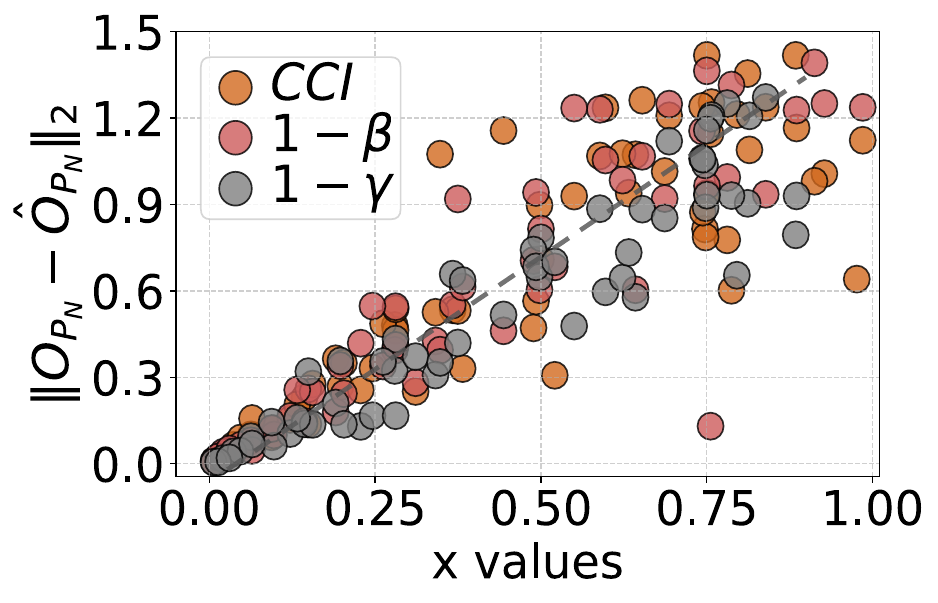}
        \caption{Output deviation with increasing $CCI$, 1-$\beta$ and 1-$\gamma$}
        \label{fig:loss_x}
    \end{minipage}
    \hfill
    \begin{minipage}[h!]{0.475\linewidth}
        \centering
        \includegraphics[width=1.0\linewidth]{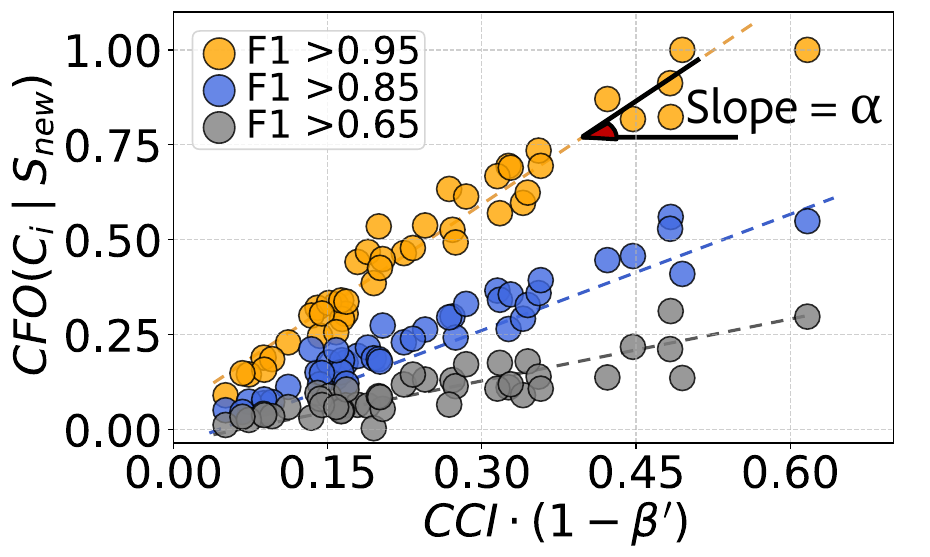}
        \caption{$CFO$ as a function of $CCI$ and $1-\beta'$ to find $\alpha$ in Eq. \ref{eq: alpha}}
        \label{fig:recomp_cfo}
    \end{minipage}
\end{figure}

\begin{figure}[t!]
    \centering
     \begin{minipage}[h!]{0.44\linewidth}
        \centering
        \includegraphics[width=1.0\linewidth]{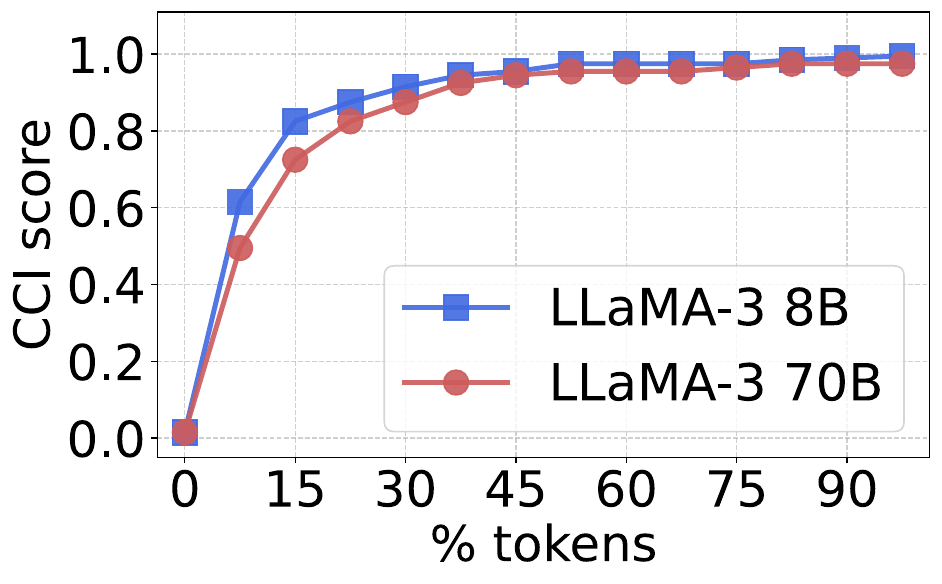}
        \caption{$CCI$ score is majorly from top recomp candidates}
        \label{fig:cci_tokens}
    \end{minipage}
    \hfill
    \begin{minipage}[h!]{0.475\linewidth}
        \centering
        \includegraphics[width=1.0\linewidth]{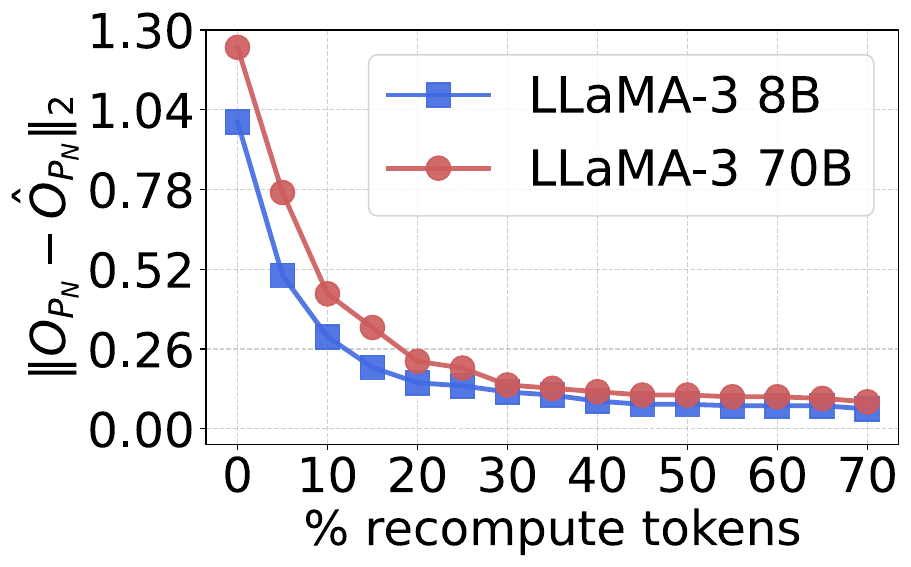}
        \caption{Output deviation decreases with higher recomputation}
        \label{fig:loss_recomp}
    \end{minipage}
\end{figure}

\subsection{Fixing {Chunk-Cache} via Recomputation} \label{sec: fixing_recomputation}  
\hl{We \textit{fix}} the \textit{chunk-caches} at runtime to make them reusable across different contexts. 
Fixing refers to recomputing only the necessary KV values to ensure the output of the reused cache closely mimics the output without any cache reuse. Fig. \ref{fig:loss_x} shows how output deviations increase with higher $CCI$ and higher $1 - \beta'$, indicating greater fixing requirements. 
$CCI$ captures the chunk’s contextual dependency, while \(1 - \beta'\) reflects prefix mismatch. 
We use this to define Cache Fix Overhead ($CFO$) for chunk $C_i$ as
\begin{align}
\label{eq: alpha}
    CFO(C_i \mid S_{new}) = \alpha \cdot CCI \cdot (1 - \beta'),
\end{align}
\hl{where $\alpha$ is a scaling hyperparameter that adjusts the recomputation factor. A higher value of $CFO$ indicates a higher fraction of the tokens in $C_i$ needs KV recomputation: $CFO=1$ for recomputing all tokens in $C_i$.} 

\noindent\textbf{Setting $\alpha$ in deployment:}
\hl{
As we lower the value of \(\alpha\), and corresponding $CFO_{\alpha}$, in expectation we would employ less recomputation per request. This might lead to corresponding quality score ($\text{F1}_\alpha$) to go down below the acceptable level ($\text{F1}_{desired}$).
We determine \(\alpha\) from a validation dataset by solving:
\begin{align}
    \alpha^* = \arg\min_{\alpha} \mathbb{E}[\text{CFO}_\alpha],\quad \text{subject to} \quad \text{F1}_\alpha \geq \text{F1}_{\text{desired}}.
\end{align}
%
Fig. \ref{fig:recomp_cfo} plots $CFO$ against $CCI\cdot(1 - \beta')$ for different $\alpha$ values and F1 scores on 2WikiMQA \cite{ho-etal-2020-constructing} dataset.
}

\subsubsection{Token Selection for Recomputation:}
\label{sec: token_selection}

\begin{figure}[t!]
    \centering
    \begin{minipage}[t]{0.42\linewidth}
        \centering
        \includegraphics[width=0.9\linewidth]
    {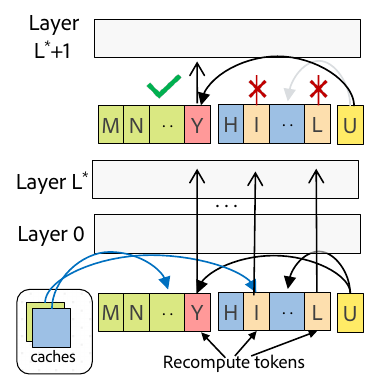}
        \caption{Algorithm selects \textit{"focused" Chunks}  across layers to reduce recomputation}
        \label{fig:focus_chunks}
    \end{minipage}%
    \hfill
    \begin{minipage}[t]{0.26\linewidth}
        \centering
        \includegraphics[width=1.0\linewidth]{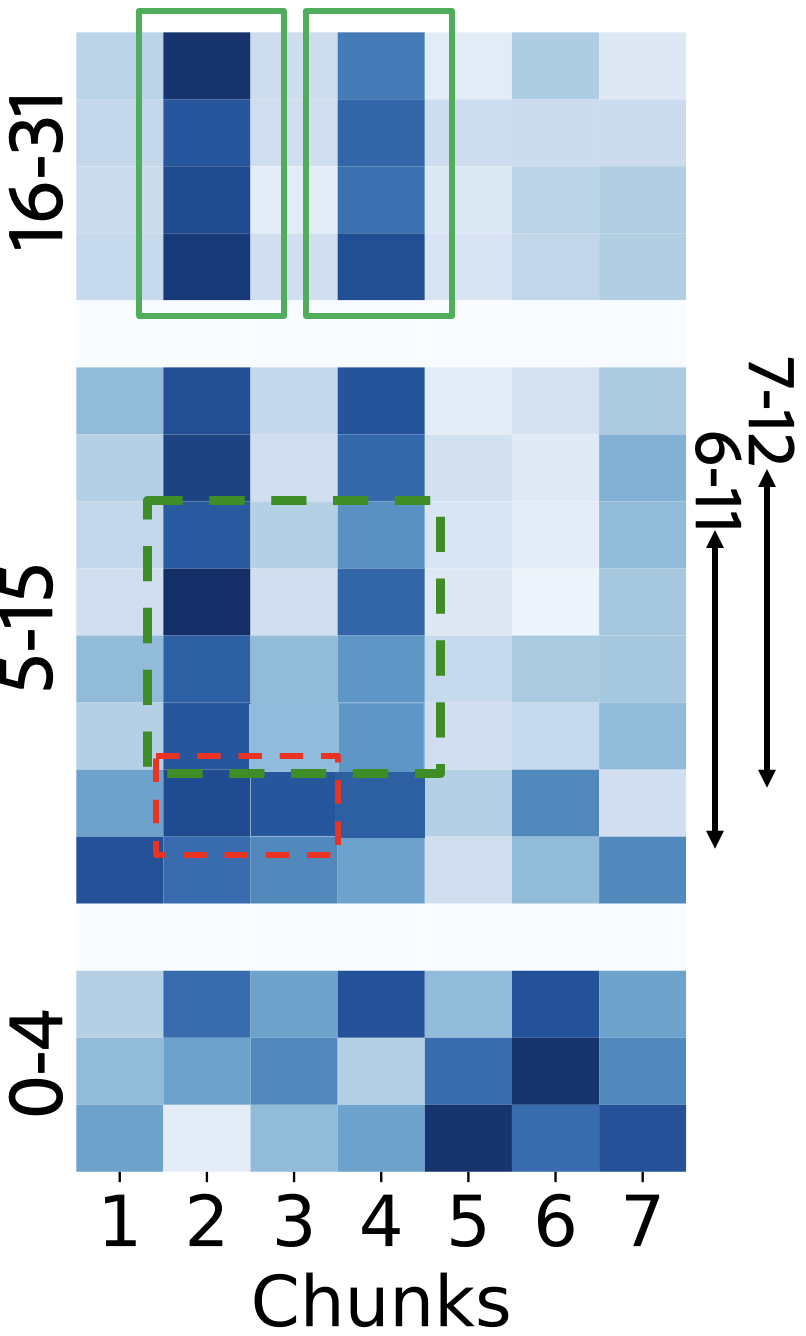}
        \caption{Question to Chunks attention across layers}
        \label{fig:layers_signal}
    \end{minipage}%
    \hfill
    \begin{minipage}[t]{0.26\linewidth}
        \centering
        \includegraphics[width=\linewidth]{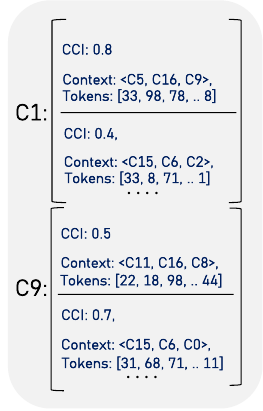}
        \caption{Metadata Store for \textit{"chunk-cache"} lookup}
        \label{fig:chunk_level_metadata}
    \end{minipage}
\end{figure}

We have observed that a small subset of tokens in a chunk significantly impacts the \(CCI\) score (Fig.~\ref{fig:cci_tokens}). Also, recomputing these critical tokens reduces output deviation (Fig. \ref{fig:loss_recomp}). Hence, to reuse a \textit{chunk-cache}, we focus on recomputing the top $N = \lceil CFO(C_i) \cdot |C_i| \rceil$ tokens with the highest inter-attention scores from prior chunks.
We select the top-N contextualized tokens for chunk $C_{i}$ as
\hl{\begin{align}
    \mathbb{T}(C_{i}) = \arg top_{N}\Big(\Big\{ \sum_{j<i} inter( C_{j}, t_k) \Big\}_{t_k \in C_i}\Big),
    \vspace{-5mm}
\end{align}}
where \hl{${inter}(t_k, C_j)$} denotes the inter-attention score between token $t_k$ in chunk \hl{$C_i$ and a prefix chunk $C_j$}. This method ensures the selection of the most contextualized tokens for recomputation.

\subsubsection{Adaptive Early \hl{Recomputation Termination}.} \label{sec: focus_selection}

In RAG pipelines, it is a standard practice to retrieve a sufficient number of chunks from a large knowledge base and utilize the LLM to filter out irrelevant content for coherent answers~\cite{wang2023learning}. In \sys, we leverage this characteristic to reduce runtime recomputation costs.

\hl{In each layer $l$, during the recomputation of selected tokens,} 
we monitor the attention between a chunk $C_i$ and the current question $U$, i.e., $inter_l(C_i, U)$, to identify the chunks that consistently receive \textit{"focused"} attention from $U$
as shown in Fig. \ref{fig:focus_chunks}. 
We find that while \hl{the inter-attention scores vary during the initial layers}, after a certain number of layers, \hl{they settle into values that can segregate the focused chunks from the others.}
Fig. \ref{fig:layers_signal}  illustrates that for approximately 80\% of queries, focused chunks can be detected between layers 10 and 15 for the \llama-3-8B model.
Consequently, we early-terminate the recomputation of tokens in the \textit{"unfocused"} chunks to minimize unnecessary computations. 

\hl{
Algorithm~\ref{algo:predict_focused} details our method for predicting focused chunks, drawing ideas from change-point detection ~\cite{aminikhanghahi2018real}.
In each a layer $l$, we first calculate the inter-attention scores w.r.t. the user question $U$ for each chunk cumulated up to layer $l$: 
\begin{align}
    cinter_l(C_i, U) = \sum_{l'=1}^l inter_{l'}(C_{i}, U) \quad \text{for all } i \in [k].
\end{align}
Then based on these cumulative scores, we segregate the high-valued chunks from the low-valued ones in an adaptive manner (lines 5-9). If this set of high-valued chunks does not change for $w$ consecutive layers, then we deem them as the focused chunks and stop the recomputation for other chunks. }

In \S\ref{sec:discussion} (Fig. \ref{fig:design_ablate}), we observe that this approach reduces token recomputation by about 55\% while maintaining similar output quality.

\subsection{Cache Variants: Retrieval and Eviction}
\label{sec:variants}

\sys maintains a data structure, \hl{as shown in Fig.~\ref{fig:chunk_level_metadata}}, for efficient lookup, retrieval, and eviction. Each \textit{chunk-cache} is identified by hashing the original chunk texts linked to the RAG vector similarity search (\S\ref{sec:intro}). This results in a \texttt{map} where chunk hashes serve as keys and lists of prefixes for each chunk are stored as values. \sys targets to store $N \times M$ \textit{chunk-cache} instances, starting with $N$ chunks (the number of keys in the \texttt{map}), each having $M$ variants. These variants help \sys recover from cases where the initial  \textit{chunk-cache} may not be optimal (e.g., excessive token recomputation due to high contextualization), while subsequent \textit{chunk-cache} variants may be more reusable for common contexts. Each variant stores the $CCI$ value and an ordered list of token indices needing recomputation. To find the best \textit{chunk-cache} for a request, \sys calculates the reusability score \(CFO = CCI \times (1 - \beta')\) (as discussed in \S~\ref{sec:chunk_reuse}) and selects the variant with the lowest score to minimize token recomputation.

For each \textit{chunk-cache} access, \sys updates its \textit{frequency-reuse} ($f_r$) as \(f_r \mathrel{+}= 1/CFO\). Consequently, \textit{chunk-caches} with higher prefix matches or less contextualization become more reusable, as indicated by increasing \(f_r\) over time. New variants are added when \sys encounters a unique chunk and prefix until it reaches $N \times M$ instances. After this, \sys periodically evicts caches with the lowest \(f_r\) to make room for more effective variants. This allows diverse configurations, from one popular chunk with \hl{$N \times M$ variants to $N \times M$} chunks, each with a single variant.

This design enables \sys to manage storage dynamically, prioritizing caches that maximize reusability while minimizing recomputation, thus reducing prefill computation. Traditional policies like LRU, LFU, or FIFO do not offer this capability. The choice of \(M\) and \(N\) is influenced by the popularity and reusability of the \textit{chunk-caches}, the RAG setting (i.e., the number of retrieved chunks), the architecture (GPU/CPU memory size and interconnects), and the deployment configuration of the LLM.

\subsection{{Chunk-Cache} Reuse Pipeline}
\label{sec:overview_pipeline}

\begin{algorithm}[t]
    \caption{Predicting Focused Chunks}
    \label{algo:predict_focused}
    \scalebox{0.9}{  
    \begin{minipage}{1.15\linewidth}
    \begin{algorithmic}[1]
    \REQUIRE \makebox[0.9\linewidth][l]{$L$: total number of layers, $w$: layer confidence window}
    \ENSURE \makebox[0.9\linewidth][l]{$F^*$: set of focused chunks, $L^*$: recomputation cut-off layer}
    
    \STATE $F_{all} \gets [ ]$, \ $cinter_i \gets 0 \quad \forall i \in [k]$
    \FOR{$l=1$ to $L$}
        \STATE $cinter_i \ +\!= \ inter_l(C_i, U) \quad \forall i \in [k]$
        \STATE $sorted\_cinter \gets \textbf{sort}([cinter_i \ |\  i \in [k]], \text{descending})$
        \STATE $diff \gets [sorted\_cinter_i - sorted\_cinter_{i+1} \mid i \in [k-1]]$
        \STATE $p_i \gets \frac{diff_i}{\sum_{i=1}^{k-1} diff_i} \quad \forall i \in [k-1]$
        \STATE $h_i \gets -\sum_{j=0}^{i} p_j \cdot \log(p_j) \quad \forall i \in [k-1]$
        \STATE $i^* \gets \textbf{argmax}([h_{i+1} - h_{i} \mid i \in [k-2]])$
        \STATE F = top $i^*$ chunks in $sorted\_cinter$
        \STATE $F_{all}$.\text{append}(F)
        \IF{$l \geq w$ \& {is\_all\_equal}($F_{all}[l-w:l]$) == 1 }
            \STATE $F^*, L^* \gets F, l$
            \STATE 
            \textbf{Break}
        \ENDIF
    \ENDFOR
    \RETURN{} $F^*,\ L^*$
    \end{algorithmic}
    \end{minipage}
    }
    \end{algorithm}

\sys implements an efficient LLM inference pipeline to minimize redundant computations in RAG by strategically reusing \textit{chunk-caches} across prefill requests.

\subsubsection{Recomputation Planning:} 
\label{sec: request_manager}

For a \hl{user query $U$}, the prefill request consists of ordered chunks \( C_1, C_2, \ldots, C_n \) provided by RAG. The system first queries the \textit{Metadata Store}, \hl{a CPU-memory-based hash-table}, to determine which chunks have their \textit{chunk-caches} available. Based on this, the chunks are then classified into two subsets: \( C_{hit} \) (with \textit{chunk-caches}) and \( C_{miss} \) (without \textit{chunk-caches}). 

It then generates an Inference Plan, designating chunks in \( C_{miss} \) for \textit{chunk-cache} computation and those in \( C_{hit} \) for \textit{chunk-cache} retrieval. It uses the metadata retrieved from the \textit{Metadata Store} to compute the \textit{Adjusted Prefix Overlap} score (\(\beta'\)) and the \textit{Chunk Context Impact} score (\(CCI\)) and then determines the \textit{Cache Fixing Overhead} (\(CFO\)) (\S\ref{sec: fixing_recomputation}). Finally, the \textit{top-N contextualized tokens} \(\mathbb{T}\), that need to be recomputed for each  \( C_{hit} \) \textit{chunk-cache} are identified.

\hl{Note that both the \textit{Metadata Store} and \texttt{vLLM}'s KV-block manager are distinct CPU-memory hash-tables. While the \textit{Metadata Store} tracks RAG chunk metadata, the KV-block hash-table maps tokens to their respective KV-blocks. Details on enabling independent chunk access without relying on prior prefixes are provided in }\S\ref{sec:impl}.

\subsubsection{\hl{Layer-wise Preloading of \textit{cache-chunks}:}}
\label{sec: preloading_kv}

\hl{
\sys uses a hierarchical caching mechanism across GPU memory, CPU memory, and SSD to expand the effective memory capacity available to store the \textit{chunk-caches}. To reduce average loading latency, the most frequently used \textit{chunk-caches} are stored in GPU, less frequently ones in CPU, and infrequent ones in SSD.
Further, \sys uses a layer-wise preloading technique to minimize cache loading delays.} 
While the GPU processes layer \(l\), the \textit{chunk-caches} for the layer \(l+1\) are concurrently loaded from host memory or SSD.
Specifically, \textsc{Cache-Craft} overlaps the loading of caches for $C_{hit}$ chunks for layer \(l+1\) with two activities: a) prefill computation of new \(C_{miss}\) chunks and b) KV recomputation of tokens in $C_{hit}$ chunks in layer \(l\). This ensures that by the time the GPU begins computing attention for layer \(l+1\), the corresponding \textit{chunk-caches} are already available in the execution buffer.

However, preloading may not fully overlap with computation if the \textit{chunk-cache} loading time exceeds the computation time for a layer, particularly when loading from SSDs. To address this, \sys reserves an HBM read buffer that allows preloading \textit{chunk-caches} for multiple layers in advance. We determine the optimal preloading depth \hl{$L_{p}$} as:
\begin{align}
    L_p = (L-1) \left( 1 - \frac{T_{prefill}}{T_{load}} \right) + 1,
    \label{eq:layerwise}
\end{align}
\hl{where $L$ is the total} number of layers, $T_{prefill}$ is prefill computation time and $T_{load}$ is KV loading time.
\hl{The goal is to} preload \(L_p\) layers such that the \textit{chunk-caches} for the remaining \((L-L_p)\) layers can be loaded within the computation time for \((L-1)\) layers.

\begin{figure}[t]
    \centering
    \includegraphics[width=0.85\linewidth]{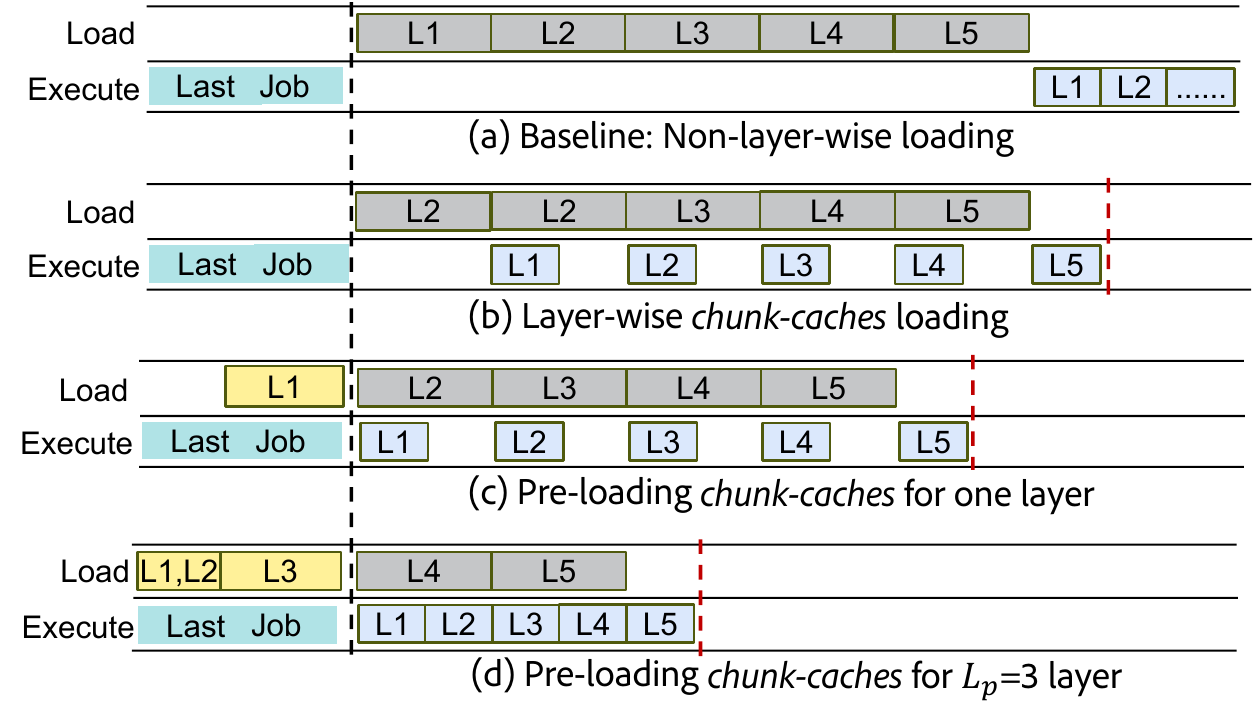}
    \caption{Layer-wise preloading of \textit{chunk-caches} into GPU memory to eliminate wait time during prefill execution.}
    \label{fig: layerwise}
\end{figure}

\hl{Algorithm \ref{alg:layerwise_preloading} shows how layer-wise preloading is implemented.}
When \(T_{load} > T_{prefill}\), preloading \(L_p\) layers minimizes wait times by eliminating layer execution gaps. If \(T_{prefill} \geq T_{load}\), preloading just one layer is sufficient due to the longer prefill time. Fig. \ref{fig: layerwise} shows an example for \(L=5\) layers with a \(T_{prefill} : T_{load}\) ratio of 1:2 where preloading \(L_p=3\) layers eliminates execution gaps.

\begin{algorithm}[t]
\small
\caption{\hl{Layer-wise Preloading of Chunk-Caches}}
\label{alg:layerwise_preloading}
\begin{algorithmic}[1]
\REQUIRE $L$: Total layers, $T_{\text{prefill}}$: Prefill time, $T_{\text{load}}$: Load time
\ENSURE Minimized execution gaps

\STATE $L_p \gets \max(1, (L-1) \cdot (1 - T_{\text{prefill}} / T_{\text{load}}) + 1)$
\FOR{$i = 1$ to $L$}
    \FOR{$j = i$ to $\min(i+L_p, L)$}
        \STATE Preload layer $j$
    \ENDFOR
    \STATE Compute layer $i$
    \STATE Release resources for layer $i$
\ENDFOR
\end{algorithmic}
\end{algorithm}

\subsubsection{Handling Partial Prefill in  LLM:}
\label{sec: inference_runner}

For each layer, \hl{the Key-Value pairs (KV)} for \textit{chunk-caches} are fetched from HBM, while the K, V, and Q are computed only for new chunks and recomputation tokens.
To support \textit{chunk-cache} reuse across contexts, \sys decouples Rotary Position Embedding (RPE) from K in KV caches. This allows dynamic RPE application during inference, adapting \textit{chunk-caches} to new positions. The system first merges the retrieved and newly computed $KV$, then applies the new RPE to the merged $K$ based on updated chunk positions, and also applies RPE to the computed Query Q.

Next, attention is computed using the newly computed Q and the merged K and V. Since Q has a different shape than K and V, a custom attention mask is required, replacing the standard triangular causal mask. As shown in Fig.~\ref{fig:rag_overview}, \textit{attention scores} (Q$XK^T$) are computed with this custom mask, multiplied by V, and passed through the feedforward network (\texttt{FFN}) to produce the layer output.

During attention computation, new \textit{chunk-caches} and inter/intra attention ($QXK^T$) for the new chunks are asynchronously saved in the background. Additionally, at every layer, attention output between question and RAG chunks ($Q_{inter}^l$) is used for \textit{Focused Chunk Selection} (\S\ref{sec: focus_selection}). Once the focused chunks are determined at layer $L^*$, recomputation for "unfocused" chunks stops.
Note that the Q for the last prefill token is always computed to generate the first decode token. After prefill, the decode phase proceeds as usual with all KV data—cached, newly computed, and recomputed.

\subsection{\hl{Hierarchical {Chunk-Cache} Management}} 
\label{sec: storage,swap,loading}

\hl{
\sys manages cache storage efficiently across GPU High Bandwidth Memory (HBM), host (CPU) memory, and SSD so that less frequent caches are moved to
further locations (from GPU HBM-memory to SSD) without deleting them when the LLM requires more GPU memory.
}

To offset the loading time of caches from non-HBM locations, \sys employs preloading techniques that start to move the caches to GPU memory, asynchronously, while requests are still in the queue.
If the caches are available in GPU memory when the request is ready to be executed, \sys uses it; otherwise, it defaults to prefill from scratch starting from input text tokens. 
This technique ensures that highly reusable chunks remain in HBM while low-reuse chunks are progressively swapped to CPU-memory, and later to SSD, before eventual eviction, if not reused. 

\hl{Using such asynchronous as well as layer-wise (\S~\ref{sec: preloading_kv}) preloading, \sys significantly reduces loading delay to make chunk-caching effective.
For example, the loading time required for 5 RAG \textit{chunk-caches} corresponding to a request in \X takes \textbf{0.03s} for CPU and \textbf{0.59s} for SSD. 
In \X a typical queue wait time is \textbf{0.32s}, allowing for preloading chunks from CPU or SSD without impacting latency significantly.
For higher loads, queue time can completely mask the loading time even from the SSD.
}

\hl{\textbf{Cache Scaling and Workload Adaptability:} In production workloads, the \textit{chunk-cache} size grows with question diversity but a small set of chunks remains crucial. For example, in \textsc{Sys-X}, 20.6\% of chunks were accessed over 1 month, with 85\% of requests hitting only 13.5\% of chunks. By 3 months, chunk accesses increased to 24\%, but 85\% of requests still accessed just 14.8\%, indicating minimal growth in the required chunk set. In this workload, these stable highly reused chunks (119 GB for \texttt{LLaMA}-3-70B) fit comfortably within the 135 GB free GPU memory budget (corresponding to LLaMa-3-70B hosted on 4, A100-80GB GPUs~\cite{9361255}, with \texttt{Tensor-Parallelism}). 
Hence, our design ensures we can handle growth in \textit{chunk-cache} size in the future. 
}

\section{Implementation}
\label{sec:impl}

\noindent\hl{\textsc{Cache-Craft} is a wrapper around \vllm ~\cite{kwon2023efficient}, built on \texttt{Xformers} ~\cite{xFormers2022} backend optimized with Triton ~\cite{Triton2021}. It enables \textit{chunk-cache} reuse for prefix and non-prefix tokens by efficiently managing positional shifts and enabling partial recomputation of prefill.}

\noindent \hl{\textbf{Chunk Storage Management:} \textsc{Cache-Craft} manages \textit{chunk-caches} by implementing a hash table at the granularity of individual RAG chunks. Unlike \texttt{vLLM}, which hashes entire prefixes pointing to the start of the KV cache (spanning multiple chunks), our approach generates independent hashes for each chunk, allowing direct access without dependence on prior context. Each chunk maps to a list of 16-token memory blocks for efficient and independent access. For optimized retrieval, the hash table stores address pointers across memory tiers, prioritizing faster tiers while allowing fallback to slower ones when necessary. Variable chunk sizes are padded to align with 16-token blocks, ensuring a consistent memory layout. Such padding causes negligible output deviation.}


\noindent \hl{\textbf{RPE Management:} To enable the reuse of \textit{chunk-caches} in arbitrary positions, \textsc{Cache-Craft} stores all cached chunks without RPE and dynamically applies corrected positional embeddings during runtime based on the current context. To efficiently manage large caches, \textsc{Cache-Craft} employs a custom CUDA kernel to remove RPE from the Keys of the KV cache after processing each request. This kernel reverses the RPE operation, \( x \cos(\theta) - y \sin(\theta) \), by applying its inverse, \( y \cos(\theta) + x \sin(\theta) \), where \( x \) and \( y \) represent the upper and lower 64-dimensional components of each token's 128-dimensional embedding and $\theta$ is the rotational angle. 
\textsc{Cache-Craft} applies relative positional encoding (RPE) to cached chunks before attention computation and removes it after decoding, ensuring reusability across varying positions. In batched inference, it optimizes RPE handling by considering shared chunk positions within the batch. For requests with differing chunk positions, RPE is integrated directly into the attention mechanism during the prefill and decoding stages. For identical positions, RPE is applied before attention and removed post-decoding. This minimizes computational overhead while ensuring correct positional embedding.}

\noindent \hl{\textbf{Selective Token Recomputation:} \textsc{Cache-Craft} modifies the Triton-based flash attention kernel to enable selective token recomputation during the prefill phase. It computes QKV values for both the scattered recomputed tokens and new question tokens. The attention kernel processes queries from these tokens, performing attention computations with the entire KV matrix and parallelizing operations over recompute and question tokens. During block-wise attention, query blocks are multiplied with prior KV blocks, adhering to causal constraints enforced by a dynamic attention mask, ensuring recompute tokens attend only to preceding tokens. After the prefill phase, we inject corrected KV values for recomputed tokens into \texttt{vLLM}'s cache for autoregressive decoding. To prevent cache corruption, the updated cache is asynchronously swapped with the original after decoding.} 

\section{Evaluation}
\label{sec:eval}

\subsection{Experimental Set up}

\subsubsection{System Configuration:}
\label{sec:eval_model_and_system}
We evaluate \sys on the \llama-3 8B and 70B models~\cite{dubey2024llama} with tensor parallelism (\texttt{TP}) of 1 and 4 respectively.
All our experiments are performed on EC2 p4de.24xlarge ~\cite{AmazonEC67:online} instances with 8 A100 GPUs ~\cite{9361255} with each having 80 GB GPU (HBM) memory. The host CPU is an Intel Xeon Platinum 8275L processor with 48 cores (96 vCPUs). The instance has 1152 GB of main memory and an 8 TB NVMe SSD with a read throughput of 16 GB/s. 
The CPU and GPUs are interconnected via PCIe 4.0 ×16, providing 64 GB/s bandwidth.

\subsubsection{Datasets and Workload:}
\label{sec:eval_workload}
We evaluate our technique with 
a real production RAG workload (\X) as well as relevant datasets following previous works\cite{bai2023longbench, jin2024ragcache}.

\begin{enumerate}[]

        \item \textbf{Real-world workloads:} \X helps users set up complex workflows for an enterprise SaaS by answering questions and prescribing steps from user manuals. It retrieves top-k=5 chunks based on the query.
        As \X creates a chunk based on the subsections of the user manual, each of the chunks can have a highly variable number of tokens. This results in a total input size 
        of 1k-20k tokens with a median of 3.3k tokens (Fig.~\ref{fig:chunks_cdf}).

        \item \textbf{Single-Hop QnA:}
        A question can be answered from a single chunk for this class of datasets. SQuAD \cite{rajpurkar2016squad} focuses on extracting answers from passages, while DROP \cite{dua2019drop} requires discrete reasoning over chunks. For multi-chunk RAG with $k=5$, we selected 200 questions and split them into 512-token chunks. 

        \item \textbf{Multi-Hop QnA:} This class of datasets requires using facts and data from multiple chunks to answer each question properly. 
        We utilize 2WikiMQA \cite{ho-etal-2020-constructing} and MuSiQue \cite{10.1162/tacl_a_00475}, which are benchmarks for evaluating complex answers across multiple documents. We sampled 200 questions. 

\item \textbf{Summarization:} We use CNN dataset \cite{nallapati2016abstractive} that generates summaries of news articles from CNN, and XSUM \cite{narayan2018don} that focuses on single-sentence summaries from BBC. 
For sampling, we split long chunks into smaller segments and randomly selected top-k=5 chunks. 
This method is applied to 40 large chunks, resulting in 200 summarization tasks.
\end{enumerate}

\noindent\textbf{Cache Warm-Up:} For every dataset, we use the first 20 queries to warm up the system and set up the caches so that we can evaluate the steady-state characteristics. 

\noindent\textbf{\hl{Cache Storage:}}
\hl{We store $ N = 100 $ chunks with $ M = 5 $ variants, requiring 0.05 TB for \texttt{LLaMA}-3-8B model. Specifically, caching 100 chunks, each containing around 1000 tokens across 5 versions, consumes $ 100 \times 1000 \times 5 \times 0.1\text{ MB (per token)} = 50 \text{ GB} $. For \texttt{LLaMA}-3-70B, this increases to 150  GB with 0.3 MB cache per token}.

\noindent \textbf{Tasks:} In the Single-Hop and Multi-Hop QnA datasets, we perform both long and short answering tasks by adjusting the mother prompt, i.e., instructing the LLM. Additionally, we generate 200 True/False questions from the original dataset chunks. For the Summarization task, we focus solely on long summaries.

\subsubsection{Evaluation Metrics: } \label{sec
}

We use two quality metrics:
{ROUGE-L F1} \cite{lin2004rouge}, which measures long-answer quality in Single/Multi-Hop and summarization tasks, and {Jaccard Similarity} \cite{ivchenko1998jaccard}, which is used for short answers and True/False questions.
We also conduct a \textit{user-study} with 250 participants to assess response correctness and quality on 2wikiMQA and SQuAD datasets based on \texttt{Yes}/\texttt{No} ratings.

Note, according to several prior studies~\cite{chen2019evaluating, lin2004looking}, a 
{ROUGE-L F1} score $\ge 0.6$ is considered \text{good}, and a score $\ge 0.8$ is considered almost indistinguishable from the original answer.
From our user study, we also analyzed this correlation and found that for answers with {ROUGE-L F1} scores $\ge 0.6$ and $\ge 0.8$, 81\% and 93\% of users have given a \texttt{YES}, respectively.
For efficiency, we measure {Recompute Savings}, {Time-to-First-Token (\texttt{TTFT}} i.e., prefill latency), {System Throughput}, and {Cost Savings}.

\subsubsection{Baselines}
\label{sec:eval_baselines}

\begin{figure*}[t]
    \centering
        \centering
        \includegraphics[width=1.0\linewidth]{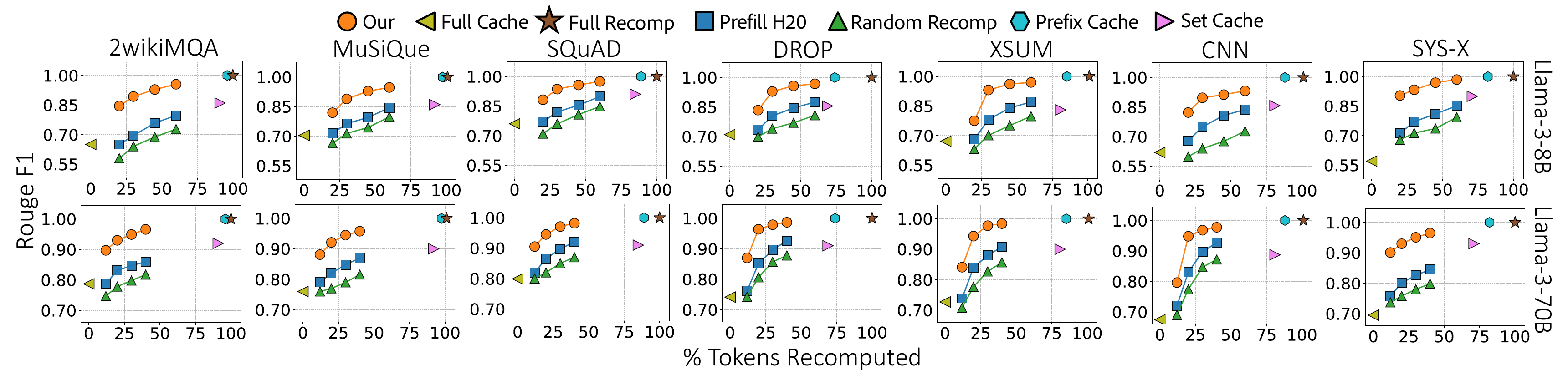}
        \caption{\hl{Rouge F1 of answer generated using Llama-3 8B and 70B on multi-hop QA, single-hop QA, text summarization, and on production \X.}}
        \label{fig:rouge_all}
\end{figure*}

\begin{figure}[t]
    \centering
        \centering
        \includegraphics[width=1.0\linewidth]{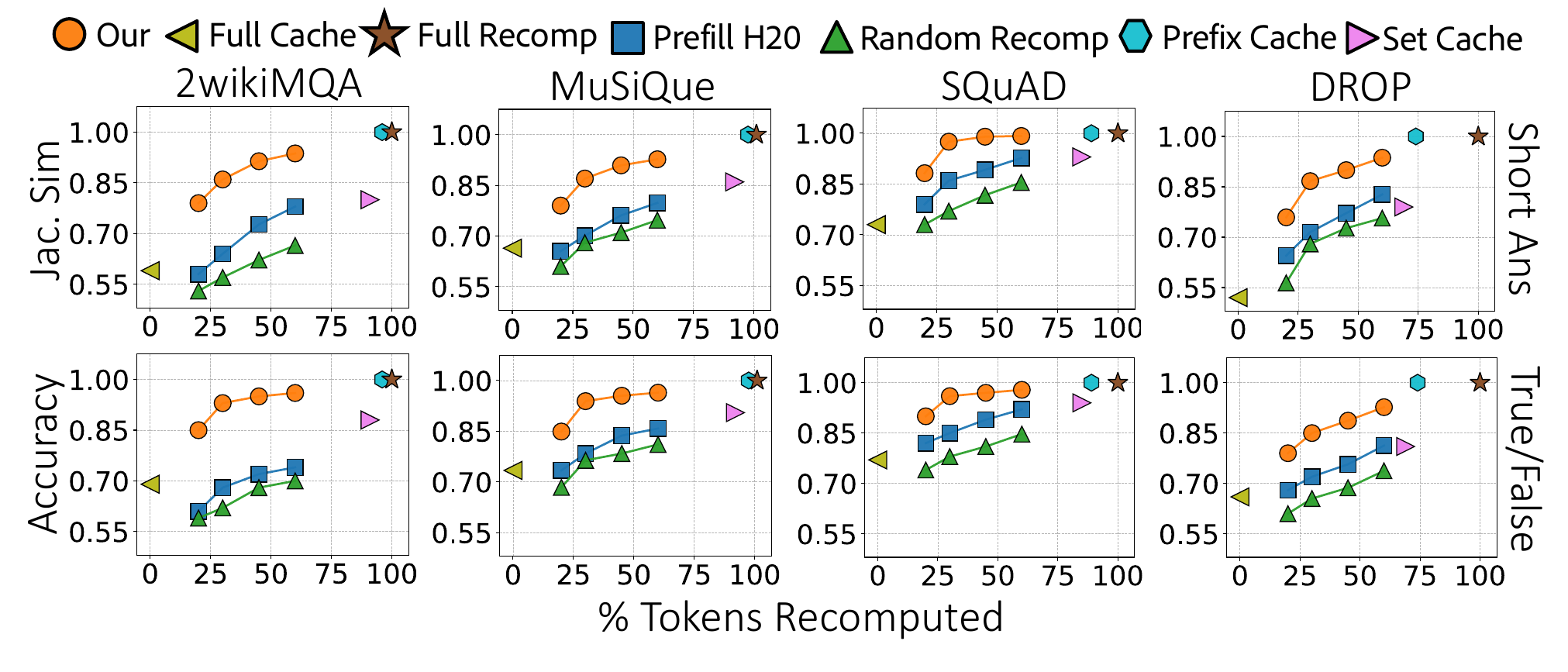}
        \caption{\hl{Jaccard Similarity and Accuracy of short answers and True/False generated using Llama-3 8B on multi-hop and single-hop datasets.}}
        \label{fig:tasks_quality_eval}
\end{figure}

We evaluate against the following baselines.
\begin{enumerate}[left=0.05em]
    \item \textbf{Prefix Matching}: We compare with two methods: (1) \textbf{\prefixcache} ~\cite{kwon2023efficient}, which reuses the KV cache based on exact prefix matches. While this approach offers perfect accuracy, it has low reuse potential. (2) \textbf{\setcache}, which modifies RPE to reorder \textit{chunk-caches} and finds the longest exact prefix match with the query. While this provides higher reuse, it has \textit{lower accuracy}.
    
    \item \textbf{Naive KV Reuse (\fullcache)}: This baseline reuses the KV cache for each chunk irrespective of the previous context, fixing only the \RPE at the new position. No recomputation is performed for the chunks.
    
    \item \textbf{Recomputation Strategies}: We also evaluate against recomputation methods: (1) \textbf{\randrecomp}, which randomly recomputes tokens within each chunk, and (2) \textbf{\ho}~\cite{zhang2024h2o}, which recomputes the most-attended tokens in the chunks. For both strategies, we maintain the same average fraction of recomputed tokens as in \sys.
    
    \item \textbf{Full Recomp (\fullrecomp)}: This \textit{\bf oracle} baseline fully recomputes all chunks for a request without utilizing any cache, providing a benchmark for optimal performance.
    
    \item \textbf{Compression Techniques}: We compare with prefill compression methods: (1) \textbf{\LLMLingua}~\cite{jiang2023llmlingua} that reduces prefill length by discarding less significant tokens using a trained model (e.g., GPT-2), and (2) \textbf{\MapReduce}~\cite{dean2008mapreduce} that summarizes context chunks for compression. The compression rates are aligned with \sys's recomputation, where 80\% compression corresponds to 20\% recompute.
\end{enumerate}

\subsection{Generation Quality with KV Chunk Reuse}
\label{sec:eval_quality}

\subsubsection{Evaluation of Recomputation Strategy}

We evaluate the recomputation strategy of \sys for Question Answering (Long and Short), True/False, and Summarization tasks using \llama-3-8B and \llama-3-70B models across multiple datasets.

 Fig.~\ref{fig:rouge_all} shows ROUGE-F1 scores, comparing \sys with baseline KV-cache reuse techniques and the original \llama generation (i.e., \fullrecomp with ROUGE score=1). 
Using \fullcache incurs no recomputation but yields low quality, ROUGE dropping to 0.65 for multi-hop QA datasets like 2wikiMQA and MuSiQue. 
In contrast, recomputing 20\% of tokens with \llama-3-8B improves the ROUGE by 30\%, and further by 42\% with 30\% recomputation. This trend is consistent across single-hop QA and summarization datasets, with $\approx$20-35\% improvements for both 8B and 70B models. \hl{Moreover, increasing recomputation to 45\% and 60\% for \textsc{LlaMa}-3-8B, and 30\% and 40\% for \textsc{LlaMa}-3-70B, further improves ROUGE scores, reaching within 1–5\% of \textsc{Full-Recomp} across all datasets.}

We also compare our contextualization-based recomputation against \randrecomp (random token selection) and \ho (high-attention token selection). Notably, random selection can lower performance even below \fullcache as it neglects the key contextual tokens and overpowers wrong tokens, which can even shadow/underpower crucial ones. \ho shows only a modest 2-10\% improvement over \fullcache but struggles with multi-hop tasks. 
\sys identifies and recomputes critical tokens distorted by prior contexts, enhancing performance and minimizing missing or incorrect facts.
Fig.~\ref{fig:tasks_quality_eval} further shows that \sys outperforms \fullcache by up to 50\% in short-QA and True/False tasks, achieving ROUGE of 0.87, compared to 0.59.

\prefixcache offers exact answers, but due to low prefix match rates, 80-95\% of tokens go through regular KV-computation, leading to very low compute savings. 
\setcache gives slightly more savings of 15-35\% by making prefixes permutation invariant with lower ROUGE due to incorrect contextualization.

To summarize, \sys offers the \textbf{\textit{best trade-off}} as its points on Fig.~\ref{fig:rouge_all} are the furthest towards the \textbf{\textit{top-left (ideal)}} corner.

\subsubsection{Comparison with Prompt Compression Techniques}

We compare \sys with established context reduction methods such as \LLMLingua~\cite{jiang2023llmlingua} and \MapReduce~\cite{dean2008mapreduce}, using datasets for multi-hop (2wikiMQA), single-hop (SQuAD), and summarization (XSUM), along with real-workload from \X, on for LLaMA-3-8B.
In Table \ref{tab:lingua_mapr_8_30}, it can be observed that with 30\% recomputation, \sys gives ROUGE-F1 scores around 0.9, which is $\approx$100\% higher than the scores for \LLMLingua (0.4) and MapReduce (0.5), for 70\% compression (i.e. comparable to 30\% recomputation).
The performance gap is due to \LLMLingua and \MapReduce's approach of discarding tokens, often losing critical information. In contrast, \sys retains all tokens by leveraging \textit{chunk-cache} reuse, ensuring no context is lost. 
Additionally, \sys selectively recomputes the most contextually impacted tokens
balancing efficiency and quality.

\subsubsection{\sys on Real Production RAG Workload:}
We evaluate \sys on production RAG workloads from \X focused on retrieval-based QA tasks where questions span multiple subsections of user manuals. As shown in Fig.~\ref{fig:rouge_all}, \sys achieves a ROUGE score of 0.87 with only 20\% token recomputation, outperforming \fullcache reuse (0.59) and other recomputation strategies by about 20-30\%. We also see that \prefixcache also saves just 18\% prefill tokens, proving ineffective. Table~\ref{tab:lingua_mapr_8_30} further compares \sys with prompt compression techniques, where \LLMLingua (0.56) and \MapReduce (0.61) score significantly lower on the \X dataset.

\begin{table}[t]
    \centering

    \begin{minipage}{0.25\textwidth}
        \centering
        \caption{ROUGE-F1 scores comparing \sys with token compression techniques for 30\% recompute tokens using \llama-3-8B}
        \label{tab:lingua_mapr_8_30}
        {\fontsize{7.75}{8.0}\selectfont
        \begin{tabular}{@{}l|c|c|c@{}}
            \toprule
            {Dataset} & {\LLMLingua} & {\MapReduce} & {\ \ Our\ \ } \\ 
            \midrule
            2wikiMQA & 32.1\% & 53.0\% & 89.3\% \\ 
            SQuAD & 45.2\% & 63.6\% & 93.6\% \\ 
            XSUM & 51.2\% & 51.6\% & 91.1\% \\ 
            \X & 56.4\% & 61.0\% & 92.0\% \\ 
            \bottomrule
        \end{tabular}
        }
    \end{minipage}
    \hspace{0.15em}
    \vline
    \hspace{0.15em}
    \begin{minipage}{0.2\textwidth}
        {\fontsize{7.75}{8.0}\selectfont
        \centering
        \caption{User study comparing \texttt{\sys} (30\% recomp) with Full Cache, Prefill H20 and Full Recompute on Llama-3-8B.}
        \label{tab:user_study}

        \begin{tabular}{@{}l|c|c@{}}
            \toprule
            & 2wikiMQA & SQuAD \\ 
            \midrule
            {\fullcache} & 29.8\% & 53.1\% \\ 
            \ho &  52.4\% & 66.8\%\\ 
            {Our} & \underline{71.2\%} & \underline{78.9\%} \\ 
            {\fullrecomp} & 76.9\% & 83.7\% \\ 
            \bottomrule
        \end{tabular}
        }
    \end{minipage}
    
\end{table}

\subsubsection{User Study:}
\label{sec: user_study_writting}

    
We conducted a user study with 250 participants with two datasets: 2wikiMQA and SQuAD.

\noindent\textbf{Task:}
We sampled 500 questions from each dataset, extracted 5 relevant chunks as the context from the datasets using vector-similarity search, and then generated answers using \hl{four methods:} (1) pass the text tokens of the chunks and the question to \llama-3-8B for a \fullrecomp to get the answers, (2) use \fullcache that simply reuses the \textit{chunk-caches} with \llama-3-8B without any recompute \hl{(3) \textsc{Prefill-H20} allows \textit{chunk-caches} reuse by recomputing the heavily attended 30\% tokens} and (4) \sys that decides which \textit{chunk-caches} to reuse and which tokens to recompute with 30\% recomputation.
Finally, each participant was presented with 15 questions, the relevant context, and answers generated by one of the methods and asked to mark Yes/No, based on the correctness and quality of the answers, when compared to the context.  
 
As illustrated in Table~\ref{tab:user_study}, \fullcache achieved significantly lower scores (30\% on 2wikiMQA, 53\% on SQuAD), whereas \sys consistently outperformed it (71\% on 2wikiMQA, 79\% on SQuAD). 
\hl{We also tested an alternative recomputation strategy, \textsc{Prefill-H20}, which showed moderate improvement (52\% on 2wikiMQA, 67\% on SQuAD), outperforming \textsc{Full-Cache} but still lagging behind \textsc{Cache-Craft}.}
It is interesting to observe that even answers from \llama-3-8B without any \textit{chunk-cache} reuse (\fullrecomp) did not get 100\% Yes, it got 77\% on 2wikiMQA, 83\% on SQuAD which is only marginally better than \sys.  
The preference for \sys was also statistically significant (\textit{p-value} < 0.05).

\begin{figure}[t]
    \centering
        \centering
        \begin{subfigure}[t]{0.48\linewidth}
            \centering
            \includegraphics[width=1.0\linewidth]{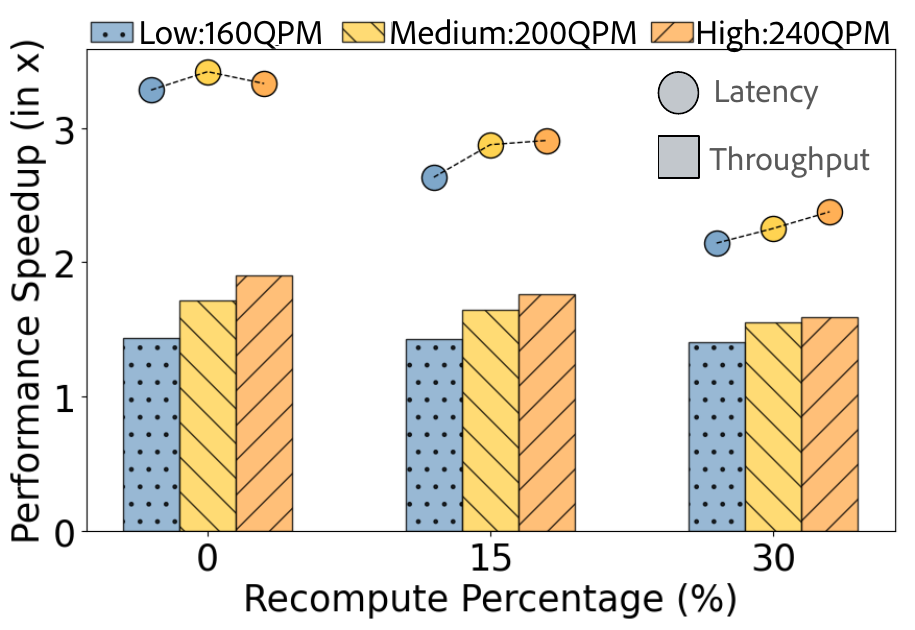}
            \caption{\hl{{\sys} performance on \llama-3-8B on 1 A100-80GB for \X }}
            \label{fig:throughput_gain}
        \end{subfigure}
        \hfill
        \begin{subfigure}[t]{0.48\linewidth}
            \centering
            \includegraphics[width=1.0\linewidth]{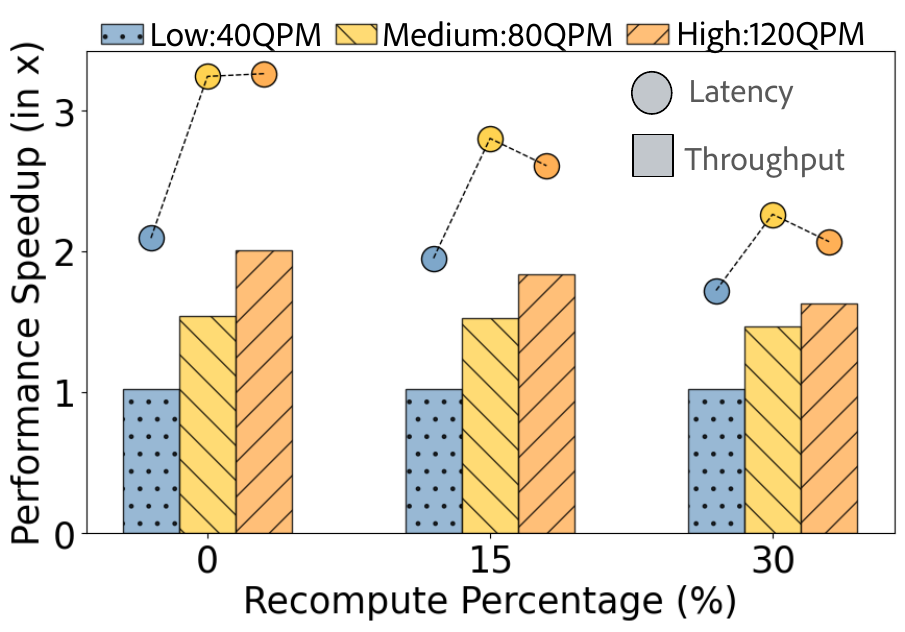}
            \caption{\hl{{\sys} performance on \llama-3-70B on 4 A100-80GB for \X}}
            \label{fig:latency_red}
        \end{subfigure}
        \caption{\hl{Throughput and overall system response latency speedup under varying computational loads for {\sys} deployed with \vllm using ORCA.}}
        \label{fig:orca_perf}
\end{figure}

\subsection{\hl{Performance Evaluation in Deployment}}
\label{sec:deploy_performance_eval}
\hl{We evaluate throughput and overall response latency under \textit{continuous batching} through \textit{ORCA}~\cite{yu2022orca}. In continuous batching, instead of waiting for all requests in a batch to complete before starting a new batch, it continuously schedules a new request when a request in the processing batch completes and slots are available. We use \X workload for both \llama-3 8B and 70B models on A100-80GB GPUs with a TP of 1 and 4, respectively. The maximum number of batched tokens in \textit{ORCA} is set to 150k tokens. The workload arrival patterns are based on public traces from \cite{proteus24,romero2020infaasmodellessmanagedinference} (which is based on Twitter traces) and proprietary data traces from Sys-X.}

\subsubsection{\hl{Throughput and Response Latency with \textit{Continuous Batching}: }} 
\hl{As shown in Fig.~\ref{fig:orca_perf}, \sys achieves up to a 1.9$\times$ speedup in throughput and a 3.3$\times$ reduction in response latency under a heavy load of 240 QPM (Queries per minute) for the \llama-3-8B model and for \llama-3-70B, it provides a 2$\times$ speedup in throughput and a 3.3$\times$ reduction in response latency under a similar heavy load of 120 QPM with no recomputation.
With 30\% token recomputation, maintaining 90\% of the base ROUGE F1 score on average, we still observe a 1.6$\times$ speedup in throughput and a 2.1$\times$ reduction in response latency for \llama-3-8B and for \llama-3-70B, the improvement is a 1.6$\times$ speedup in throughput and a 2$\times$ reduction in response latency under high load. Notably, a 30\% recomputation level for \llama-3-70B is sufficient to ensure a minimum of 90\% of the base ROUGE F1 score. 
The overall response latency reduction for \llama-3-70B under high load with 30\% recomputation is 2.07$\times$, compared to 2.26$\times$ under medium load. This difference arises due to the significantly higher wait time overhead at high load ($\approx$7.15s on average) compared to medium load ($\approx$2.15s on average). However, when excluding request wait time, the latency reduction at high load is even more pronounced (3.22$\times$ compared to 2.64$\times$).}

\hl{
\subsubsection{Preloding in Hierarchical Caching:}
We evaluate how asynchronous (\S\ref{sec: storage,swap,loading}) and layer-wise (\S\ref{sec: preloading_kv}) preloading help in hierarchical caching. 
In Fig.~\ref{fig:hier_cache} we show the timings to load the cache from CPU and SSD to GPU-memory when we start loading after the request reaches the head of the queue (\texttt{Sync}), when asynchronous preloading starts when the request is in the queue (\texttt{Async}), and when layer-wise preloading is used with \texttt{Async} (\texttt{Layer}). Cache loading takes 0.03s from the CPU and 0.59s from the SSD, adding significant overhead. With an average queue wait time of 0.32s (for \X), asynchronous preloading eliminates CPU overhead and reduces SSD overhead to 0.27s, as 0.32s overlaps with queue time.
Through layer-wise preloading, loading only the first 24 (out of a total of 32) layers in advance further reduces SSD overhead to 0.12s. CPU loading overhead is 0s for \texttt{Async} and \texttt{Layer} because loading time is already less than queue time. We compare our effective prefill time with the time taken to recompute the entire context in the fastest scenario, i.e. when the system is idle. With layer-wise preloading, our effective prefill time is shorter for both CPU and SSD. Note that for caches stored in GPU memory, there is no additional overhead for loading. In all three cases, the cache is brought to the GPU, and the time required for any retrieval from GPU memory for processing is already included in the TTFT.}

\subsection{\hl{Controlled Evaluations for TTFT}}
\label{sec:eval_performance}

We evaluate the \hl{TTFT Latency} of the \vllm implementation of \sys on the \X production workload. We also evaluate across various generation settings to ensure generalization to different models and datasets.  We compare performance for the setting for which \sys and the baselines achieve the same quality of ROUGE F1 score of 0.85. 
For \sys, \LLMLingua, and \ho, this corresponds to 25\% recomputation, 25\% compression, and 60\% recomputation, respectively. 
For \prefixcache, we copious scope by setting that 60\% of the prefill tokens will have a prefix match. Note that this is significantly higher than what (18\%) we observed for production workloads \X.

\begin{figure}[t]
    \centering
    \begin{subfigure}[t]{0.49\linewidth}
        \centering
        \includegraphics[width=0.925\linewidth]{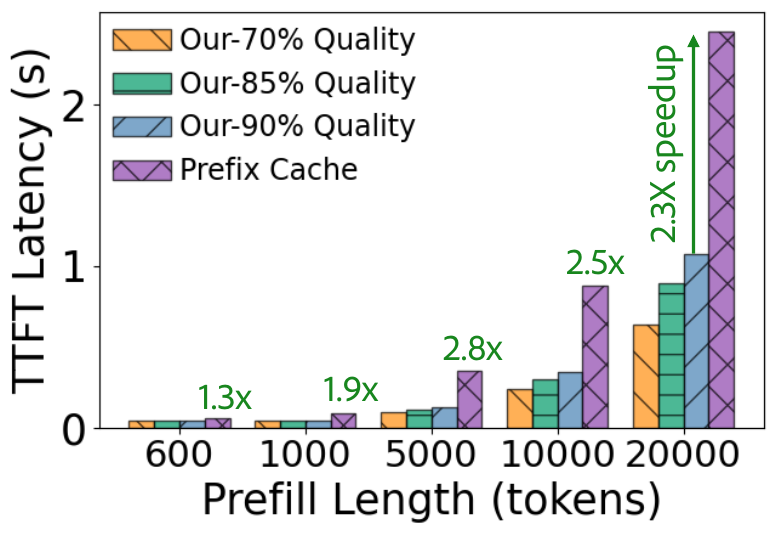}
        \caption{TTFT for \sys on \X data across different text quality}
        \label{fig:perf_compare_diff_qual}
    \end{subfigure}
    \hfill
    \begin{subfigure}[t]{0.49\linewidth}
        \centering
        \includegraphics[width=0.85\linewidth]{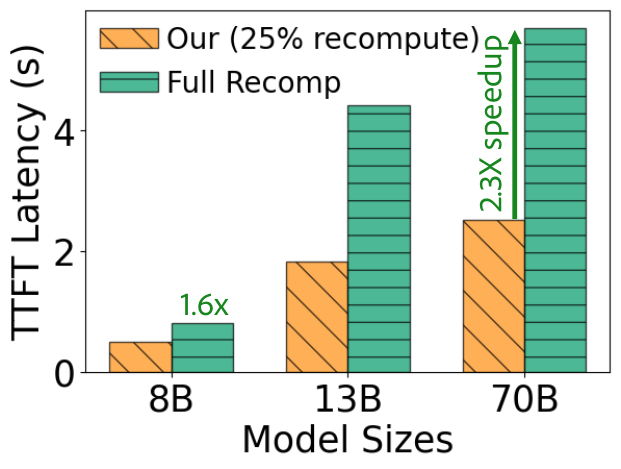}
        \caption{TTFT for different model sizes (prefill tokens 8192, batch size 4, TP 4)}
        \label{fig:model_sizes_perf}
    \end{subfigure}
    \begin{subfigure}[t]{0.49\linewidth}
        \centering
        \includegraphics[width=0.925\linewidth]{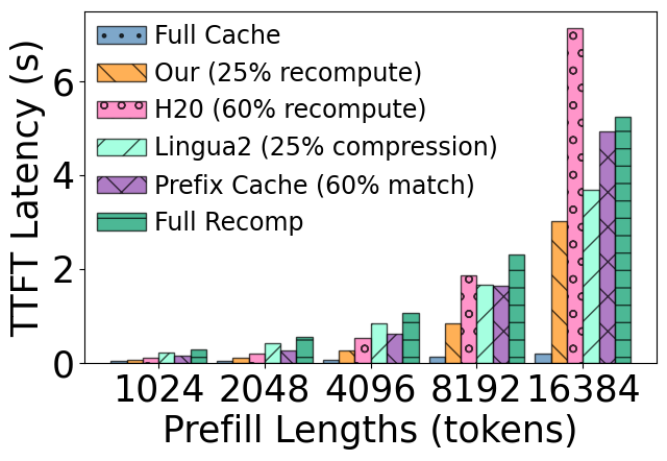}
        \caption{Baseline comparison across prefill lengths (batch size 4, TP 1)}
        \label{fig:seq_len_perf}
    \end{subfigure}
      \hfill
    \begin{subfigure}[t]{0.49\linewidth}
        \centering
        \includegraphics[width=0.925\linewidth]{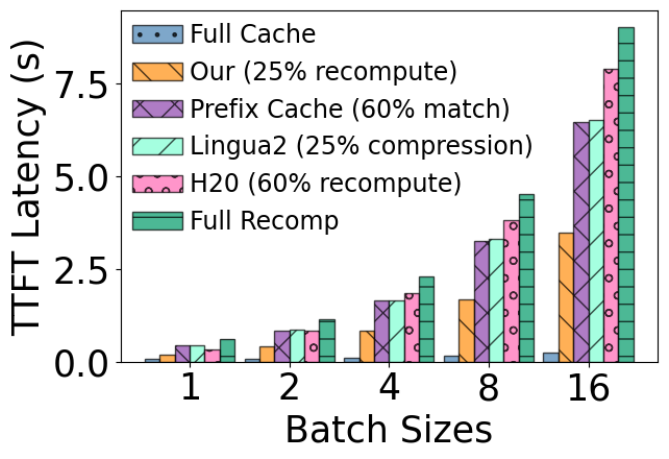}
        \caption{Baseline comparison across batch sizes (prefill tokens 8192, TP 1)}
        \label{fig:batch_size_perf}
    \end{subfigure}
    \\
    \caption{Performance evaluation of \sys on \llama-3-8B (except sizes)}
    \label{fig:2x2_subfigs}
\end{figure}

\subsubsection{Performance of \sys on \X: } In Fig. \ref{fig:perf_compare_diff_qual}, we compare the performance of \sys against Prefix Cache for requests from \X across sequence lengths. The range of sequences received by the system varies from 600 to 20000 tokens averaging around 5000 tokens. We observe about 2.5$\times$ speedup for \llama-3 8B in TTFT latency over Prefix Cache by recomputing 39\% of tokens while maintaining 90\% of the original quality. This is because on average only 18\% exact prefix match occurs for the requests received in the system rendering \prefixcache ineffective.

\begin{figure}[t]
    \centering
    \includegraphics[width=0.95\linewidth]{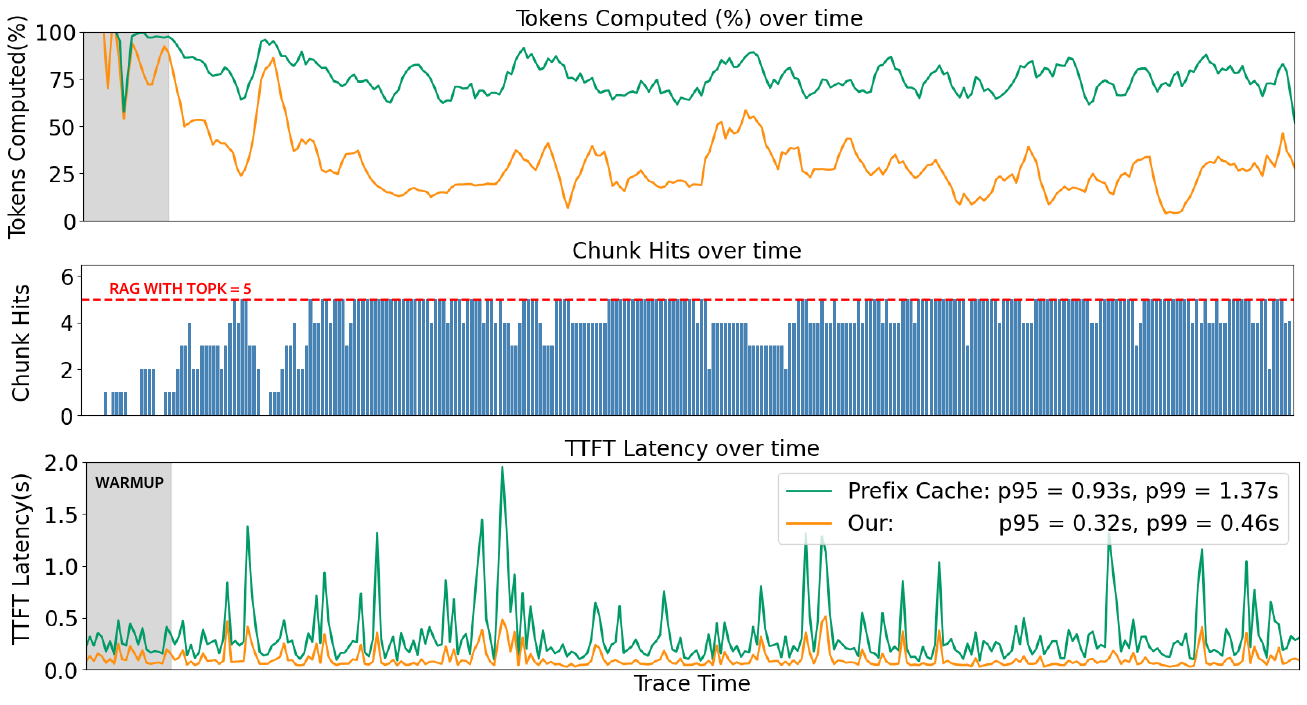}
    \caption{Evaluation of \sys on \X trace on \llama-3-8B}
    \label{fig:eval_sysx}
\end{figure}

In Fig. \ref{fig:eval_sysx}, we show TTFT latencies of each request from a trace of the requests received overtime by \X. 
We indicate the warm-up period on the left.
In the bottom plot, we observe that as \sys can keep the TTFT spikes significantly lower than vanilla \llama with \prefixcache. \sys provides 3$\times$ reduction in the 99th percentile TTFT latency. 
Note, the spikes in TTFT are due to the fact that text chunks in \X are subsections of the user manuals and are unequal in size.
However, note that when prefill lengths are high (leading to spikes), \sys comes can reduce TTFT significantly as by reusing \textit{chunk-caches}, it avoids quadratic computational complexity (\S\ref{sec: background}). 
In the top plot of Fig. \ref{fig:eval_sysx},
we also observe a consistent reduction in token computation indicating higher reusability of cached chunks compared to \prefixcache. This trend is further supported by the increasing chunk hit rate achieved by our system. This results in a 51\% average reduction in token computation compared to \prefixcache.
In the middle plot of Fig. \ref{fig:eval_sysx}, we show how many chunks for the top-k=5 retrieval in \X was a hit in the \textit{chunk-cache}. It can be observed for a large number of requests, all the necessary chunks were already in the cache --- leading to a \textit{hit-rate of 5 out of 5}.

\begin{figure}[t]
    \centering
    \includegraphics[width=0.965\linewidth, height=0.1\linewidth]{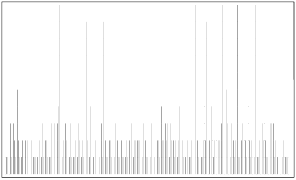}
    \caption{Snapshot of $MXN$ Cache-Store for \X trace at the end}
    \label{fig:snapshot_data}
\end{figure}

\noindent\textbf{Cache-store Characteristics:} Fig.~\ref{fig:snapshot_data} illustrates the cache-store state at the trace's end for \X. The X-axis represents the number of unique \textit{chunk-caches} (186), and the Y-axis indicates how many variants were created for each chunk (up to 11 for some). As detailed in \S~\ref{sec:variants}, \sys dynamically configures cache storage based on chunk popularity and reuse.

\subsubsection{\hl{Impact of Model Size, Prefill Length \& Batch Size}: }
Here we measure how \sys reduces TTFT latency compared to \fullrecomp across different model sizes of \llama, across different prefill-lengths, and batch sizes.
As shown in Fig. \ref{fig:model_sizes_perf}, \sys becomes more effective in reducing TTFT latency as the model size increases. This improvement is due to reduction of the number of tokens computed by \sys in each attention layer, with the gains increasing as the number of layers grows for larger models.
\sys reduces latency by 1.6$\times$ and 2.3$\times$ compared to \fullrecomp for \llama-3-8B and \llama-3-70B, respectively and with batch-size$=4$ and sequence length of 8192 tokens. 

In Fig. \ref{fig:seq_len_perf} we compare the TTFT latency against baselines, \prefixcache, \LLMLingua, and \ho, across varying sequence lengths for \llama-3-8B with batch-size$=4$. We can see \sys outperforms all baselines across different sequence lengths and for 16k length it is 1.7$\times$ faster than \fullrecomp. In Fig. \ref{fig:batch_size_perf} we show for \llama-3-8B as we increase batch-size, \sys is more effective in controlling 
TTFT latency increases than all other methods because it requires much less recomputation to maintain quality.  



\begin{figure*}[t]
    \centering
    \begin{minipage}[t]{0.21\linewidth}
        \centering
        \includegraphics[width=1.0\linewidth]{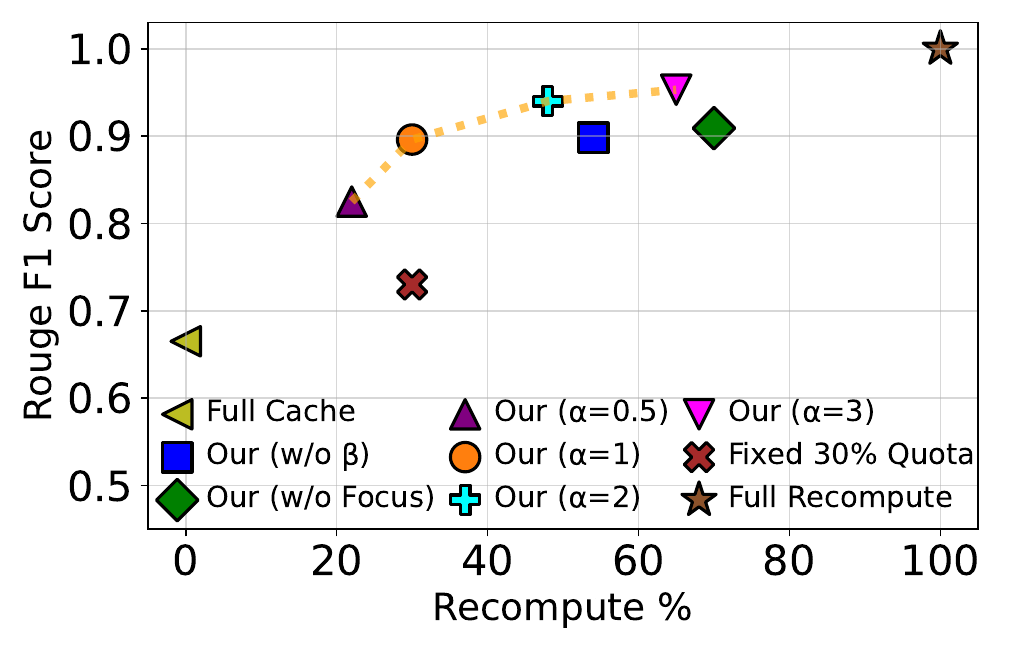}
        \caption{Design elements on Rouge F1 vs. recompute tokens for 2wikiMQA with \llama-3-8B and 30\% recomp. The dotted line connects different $\alpha$. $\alpha$=1 w/o ($\beta$ and focus)}
        \label{fig:design_ablate}
    \end{minipage}\hspace{0.25em}\vline\hspace{0.25em}
    \begin{minipage}[t]{0.21\linewidth}
        \centering
        \includegraphics[width=1.0\linewidth]{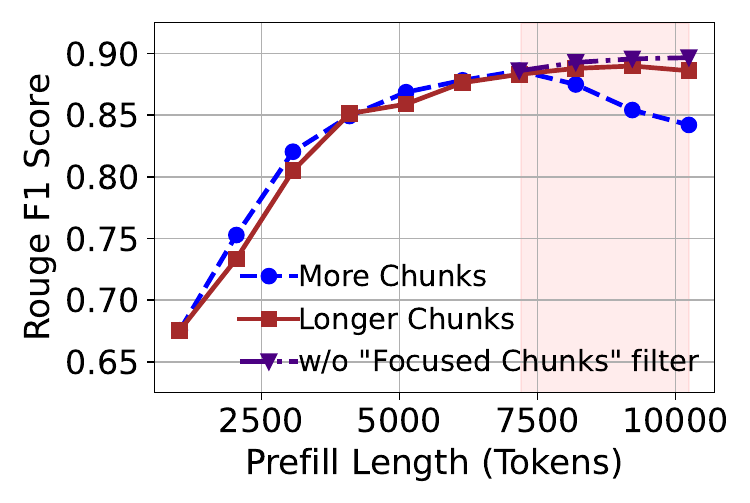}
        \caption{ROUGE-F1 quality for context length, measured by varying chunk sizes (brown) and number of chunks (blue), with \llama-3-8B at 30\% recomputation.}
        \label{fig:chunk_length_ablate}
\end{minipage}\hspace{0.25em}\vline\hspace{0.25em}
    \begin{minipage}[t]{0.21\linewidth}
        \centering
        \includegraphics[width=1.0\linewidth]{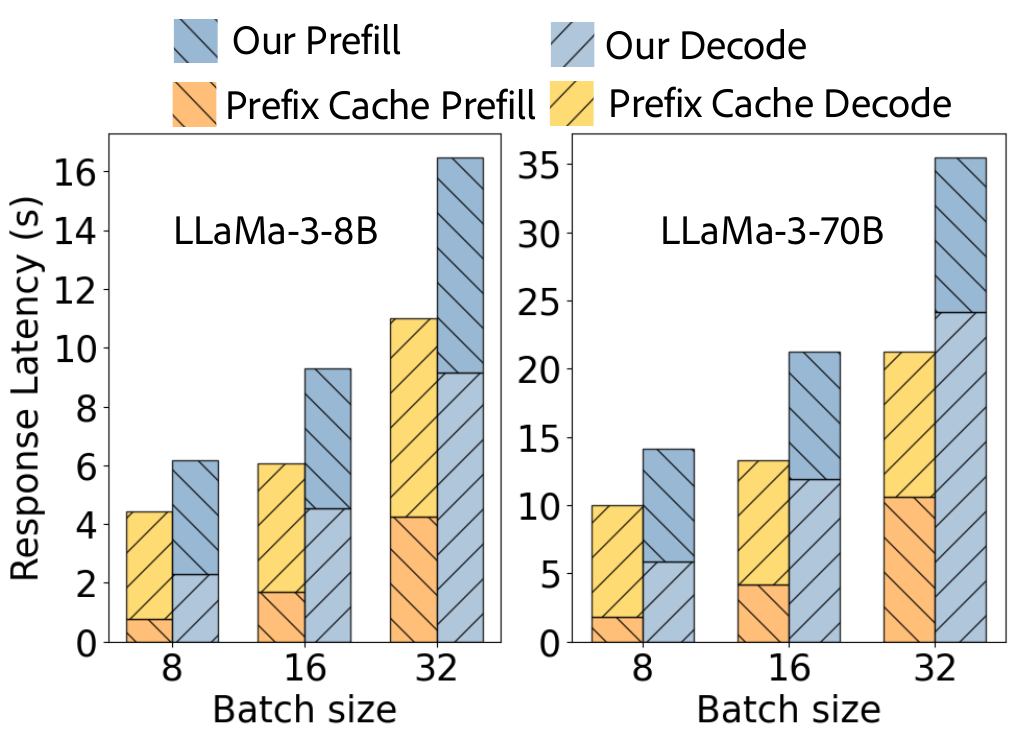}
         \caption{\hl{Prefill and decode latencies across batch size for \llama-3 8B and 70B/ Prefill takes up an increasing proportion of total time for larger batch sizes.}}
        \label{fig:batch_times_serve}
    \end{minipage}\hspace{0.25em}\vline\hspace{0.25em}
    \begin{minipage}[t]{0.165\linewidth}
        \centering
        \includegraphics[width=1.0\linewidth]{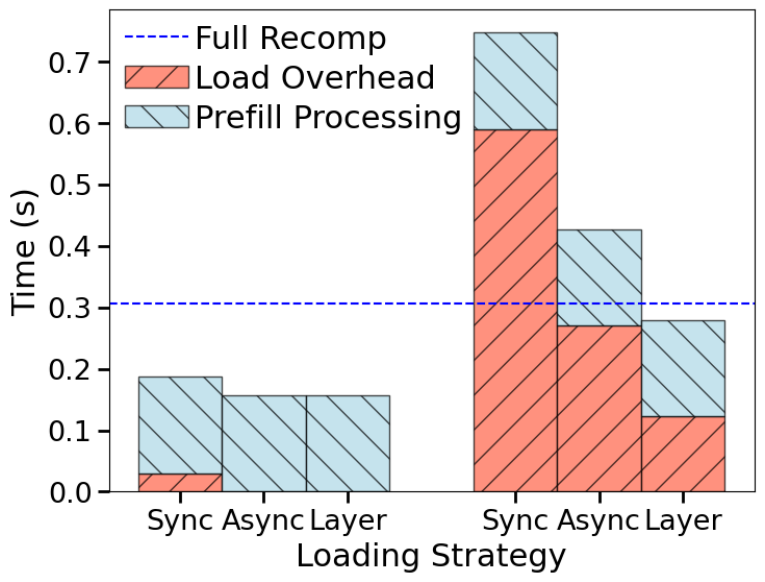}
        \caption{\hl{Cache loading overhead in \sys from different memory hierarchies using \llama-3-8B with and w/o preloading.}}
        \label{fig:hier_cache}
    \end{minipage}\hspace{0.25em}\vline\hspace{0.25em}
    \begin{minipage}[t]{0.125\linewidth}
    \vspace{-6.5em}
        \centering
         \renewcommand{\arraystretch}{0.85} 
        {\fontsize{7.1}{7.75}\selectfont
         \begin{tabular}{@{}c|c|c@{}}
        \toprule 
        {RPE} & {Causal} & {Rouge} \\ 
        \midrule
        \textcolor{red!80!black}{$\times$} & \textcolor{red!80!black}{$\times$} & 0.15 \\ 
        \textcolor{red!80!black}{$\times$} & \textcolor{green!60!black}{$\checkmark$} & 0.22 \\ 
        \textcolor{green!60!black}{$\checkmark$} & \textcolor{red!80!black}{$\times$} & 0.66  \\ 
        \textcolor{green!60!black}{$\checkmark$} & \textcolor{green!60!black}{$\checkmark$} & \textbf{0.89} \\ 
        \bottomrule
    \end{tabular}
    \vspace{0.5em}
    \captionof{table}{Impact of fixing only RPE/Causality and both RPE $+$ Causality for \textit{chunk-cache} reuse for 2WikiMQA}
    \label{tab:rope_causal_ablate1}
    }
    \end{minipage} 
    \vspace{-0.2em} 
\end{figure*}

\section{\hl{Ablations} 
 and Discussions}
\label{sec:discussion}

\textbf{Design Components in \sys: } 
The ablation study in Fig. \ref{fig:design_ablate}  on 2wikiMQA using \llama-3-8B highlights the impact of various design elements in \sys. 
We obtain a baseline score of 0.665 from Full KV Cache reuse with fixed RPE. 

Removing components of our recomputation logic-specifically, $\beta$, Cache Context Index ($CCI$), and focus chunking—provides insights into performance dynamics. Removing $\beta$ increases recomputation to 54\% without improving quality, emphasizing its role in minimizing unnecessary recomputation for well-matched chunks. Disabling focus chunking similarly raises recomputation to 70\% with minimal quality gains, underscoring the importance of both $\beta$ and focus in optimizing recompute efficiency. Additionally, when fixed recomputation is applied without $CCI$ (via random selection), 
quality declines dramatically to a ROUGE score of 0.73.

We also explore varying $\alpha$ values from 0.5 to 3, revealing an increasing quality trend: 0.825 for $\alpha=0.5$, 0.896 for $\alpha=1$, 0.94 for $\alpha=2$, and 0.953 for $\alpha=3$. However, this trend indicates diminishing returns as recomputation increases, highlighting a saturation point.

    

    

\textbf{Context Size (Number of Chunks vs. Chunk Size):} We analyze quality (ROUGE F1) trends with context lengths by varying chunk sizes (brown line) and the number of chunks (blue line) using \llama-80B with 30\% recomputation, as shown in Fig.~\ref{fig:chunk_length_ablate}. The brown line demonstrates that quality consistently increases with larger chunk sizes, stabilizing around 0.92, which underscores the reliability of our recomputation logic for longer contexts. The blue line, representing quality with more chunks, exhibits a similar upward trend but slightly declines after saturation (approximately 0.91). This drop, highlighted in red on the plot, indicates that focus chunk selection becomes less effective with too many chunks. Notably, when the "\textit{focused} chunks" filter is disabled (indigo line), quality remains stable, suggesting that the decline is attributed to the error from the "\textit{focused} chunks" selection mechanism.

\hl{\textbf{Why caching chunks is acceptable in practice?}}
\hl{Our evaluations in Fig.~\ref{fig:rouge_all} and the qualitative user study in Table~\ref{tab:user_study} demonstrate that recomputing attention scores for only 30\% of carefully selected tokens while using precomputed caches for the rest achieves 93\% of the user score attained by full attention computation while significantly improving performance. This is driven by two observations:}

\hl{\noindent\textbf{(1)} Retrieved RAG chunks, such as document sections in \X, typically exhibit low inter-dependencies, as attention scores decline with token distance. For large chunks ($>$883 tokens) in \X, the intra-attention is \textbf{2.18x} inter-attention on average with only \textbf{23\%} of chunks being highly contextualized. To address such chunks, \sys selectively recomputes KV caches for a few contextualized tokens, producing outputs close to ideal (\S\ref{sec: fixing_recomputation}). Fig.~\ref{fig:rouge_all} shows \sys offers the best trade-off between recomputation budget and quality, outperforming all baselines with a ROUGE F1 score of \textbf{0.87} using 20\% recomputation for \X~ while a threshold of \textbf{0.7} is considered good for semantic similarity \cite{li-etal-2024-traq}. This is also supported by our user study results which show \textbf{78.9\%} user
acceptance for \sys versus \textbf{83.7\%} for vanilla \vllm 
with exact computation using LLaMa-3-8B (Table~\ref{tab:user_study}).
}

\hl{\noindent\textbf{(2)} 
TTFT metric is critical in production. Current methods like prefix-caching suffer under heavy load, with TTFT reaching \textbf{35s} for LLaMa-3-70B on 4 A100 GPUs (Fig.~\ref{fig:batch_times_serve}). The proportion of TTFT in overall response latency increases with the higher system-load. \sys reduces TTFT latency by independent caching of prior context. Unlike prefix caching, which has low hit rates and high memory overhead (Fig.~\ref{fig:chunks_cdf}), \sys stores chunks independently, allowing to store more chunks in HBM and achieving higher cache hit rates by reusing chunks in different combinations.}

\textbf{Approximation (Position vs. Causal):} 
In Table~\ref{tab:rope_causal_ablate1}, we observe that reusing caches from non-prefix chunks significantly degrades performance, resulting in a ROUGE F1 score of 0.15 when neither RPE nor causality is fixed. Fixing causality without adjusting RPE yields minimal improvement (0.22) while optimizing RPE alone achieves 0.665, which serves as the \fullcache baseline. Notably, \sys achieves a ROUGE score of 0.896 with just 30\% recomputation, demonstrating that correct positional encoding combined with selective recomputation can effectively approximate the benefits of full recomputation, which scores 1.0.

\section{Related Works}
\label{sec:related_works}

\noindent\textbf{LLM Serving Efficiency:} Multiple works looked into achieving service level objectives (SLOs) at scale\cite{crankshaw2020inferline,crankshaw2017clipper,gujarati2020serving,shen2019nexus}. 
The  works in \cite{dao2022flashattention,kwon2023efficient,shi2023welder} looked into optimizations of memory, while \cite{han2022microsecond,zhang2023shepherd,yu2022orca} have explored parallelism and batching. 
\cite{dettmers2022gpt3,kwon2022fast,frantar2023sparsegpt} aimed to improve the KV computations, primarily using the model's sparsity. 

\noindent\textbf{Context Compression and KV Cache Reduction}: Improving decode speed of LLMs is a widely studied field. Several system-level techniques to optimize the prefill have been adopted \cite{qu2022dota,guo2023olive,ham2021elsa}.  
These works primarily aim to reduce the size of the KV cache during generation. Works like \cite{dong2024get,jiang2023llmlingua, jiang2023longllmlingua} focused on compressing the context length to optimize the prefill. 
Some other similar works \cite{liu2024scissorhands,zhang2024h2o,adnan2024keyformer} drop unimportant tokens while a few modify attention or use gist tokens to achieve KV reduction \cite{yan2021attention,mu2024learning}. Another set of works that aim at KV reuse has enabled increased decoding speed by saving redundant computation. Most of these assume prefix-match \cite{jin2024ragcache,liu2024optimizing,liu2023cachegen,zheng2023efficiently,gao2024attentionstore}, which is impractical for RAG-systems. 
Although \textit{Prompt Cache}\cite{gim2024prompt} enables KV cache reuse at various positions, it struggles to maintain satisfactory generation quality due to inaccuracies in positional encoding and a lack of consideration for cross-attention.
\sys enables efficient KV-cache reuse for RAG without compromising quality. 

\noindent\textbf{KV Cache Quantization and Management:} Quantization of KV-cache reduces computation while maintaining generation quality \cite{hooper2024kvquant,dong2024qaq}. Some works address fitting large KV caches in limited memory \cite{lee2024infinigen,kwon2023efficient,wang2023catalyst,wang2020put,ren2017slimdb}.
\vllm \cite{kwon2023efficient} reduces KV cache due to fragmentation.
\textit{Prompt Cache}~\cite{gim2024prompt} reuses KV-caches at different positions but relies on a rigid prompt structure, leading to poor quality when the structure changes.
These orthogonal techniques can complement \sys for additional efficiency.

\noindent\textbf{\hl{Approximation in Systems:}} 
\hl{Controlled approximation in KV-cache computation is inspired by prior works that used approximation techniques in image generation \cite{agarwal2024approximate, li2024distrifusion, ma2024deepcache,lu2024recon}, data analytics \cite{garofalakis2001approximate, park2018verdictdb, agarwal2023fast, ahmad2024scaleviz}, and video analytics \cite{xu2018videochef, zhang2017live}.}

\section{Conclusion}

We introduced \sys, a system that efficiently manages precomputed states corresponding to the chunks of the knowledge base for RAG. 
We presented several in-depth analyses of real production workloads showing several interesting characteristics of RAG-systems that show significant opportunities for \textit{chunk-cache} reuse but also highlight the technical challenges. 
With our novel technique for identifying reusable chunks, selective recompute, and cache management policies, we show that \sys can provide a significant speed-up without compromising the quality for real production workloads as well as popular RAG datasets.

\newpage

\bibliographystyle{ACM-Reference-Format}
\bibliography{main}

\end{document}